\documentclass[useAMS,usenatbib,usegraphicx]{mn2e}
\usepackage{natbib}
\usepackage{graphicx}
\usepackage{hyperref}
\usepackage{aas-macros}
\usepackage[labelfont=bf]{caption}
\usepackage{units}
\usepackage{mathpazo}
\usepackage{amsmath}
\usepackage{amssymb}
\usepackage{color}
\usepackage[T1]{fontenc} 
\usepackage{aecompl}

\newcommand{\mbh}{M_{\bullet}}

\newcommand{\msun}{{\rm M}_{\odot}}

\newcommand{\lsun}{{\rm L}_{\odot}}

\title[Erosion of Globular Cluster Systems]
{Erosion of Globular Cluster Systems: The Influence of Radial Anisotropy, Central Black Holes and Dynamical Friction}

\author[Brockamp et al.]{M. Brockamp$^{1,2}$\thanks{E-mail: \mbox{brockamp@astro.uni-bonn.de} (MB); \mbox{akuepper@astro.columbia.edu} (AHWK);
\mbox{ithies@astro.uni-bonn.de} (IT); \mbox{h.baumgardt@uq.edu.au} (HB); \mbox{pavel@astro.uni-bonn.de} (PK)},
A.H.W. K\"upper$^{3,4}$, I. Thies$^{1}$, H. Baumgardt$^{5}$, P. Kroupa$^{6}$\\
$^{1}$Argelander-Institut f\"ur Astronomie, Universit\"at Bonn, Auf dem H\"ugel 71, D-53121 Bonn, Germany\\ 
$^{2}$Member of the International Max Planck Research School (IMPRS) for Astronomy and Astrophysics at the University of Bonn and Cologne\\ 
$^{3}$Department of Astronomy, Columbia University, 550 West 120th Street, New York, NY 10027, USA\\
$^{4}$Hubble Fellow\\
$^{5}$School of Mathematics and Physics, University of Queensland, St. Lucia, QLD 4072, Australia\\
$^6$Helmholtz-Institut f\"ur Strahlen- und Kernphysik (HISKP), University of Bonn, Nussallee 14-16, 
D-53115 Bonn, Germany\\}

\begin{document}

\date{}

\pagerange{\pageref{firstpage}--\pageref{lastpage}} \pubyear{201x}

\maketitle

\label{firstpage}

\begin{abstract}
We present the adaptable \textsc{Muesli} code for investigating dynamics and erosion processes of globular clusters (GCs) in galaxies. \textsc{Muesli} follows
the orbits of individual clusters and applies internal and external dissolution processes to them. Orbit integration is based on the
self-consistent field method in combination with a time-transformed leapfrog scheme, allowing us to handle velocity-dependent forces
like triaxial dynamical friction.

In a first application, the erosion of globular cluster systems (GCSs) in elliptical galaxies is investigated. Observations show that massive
ellipticals have rich, radially extended GCSs, while some compact dwarf ellipticals contain no GCs at all.

For several representative examples, spanning the full mass scale of observed elliptical galaxies, we quantify the influence of radial
anisotropy, galactic density profiles, SMBHs, and dynamical friction on the GC erosion rate. We find that GC number density profiles are
centrally flattened in less than a Hubble time, naturally explaining observed cored GC distributions. The erosion rate depends primarily
on a galaxy's mass, half-mass radius and radial anisotropy. The fraction of eroded GCs is nearly 100\% in compact, M~32 like galaxies
and lowest in extended and massive galaxies. Finally, we uncover the existence of a violent \textit{tidal disruption dominated phase} which
is important for the rapid build-up of halo stars.
\end{abstract}


\section{Introduction}\label{sec:intro}
Globular clusters (GCs) are among the oldest objects in galaxies. They can provide a wealth of information on the formation and evolution histories
of their host galaxies as well as on cosmological structure formation \citep{1978ApJ...225..357S, 2010MNRAS.406.2000M, 2013ApJ...772...82H}. This paper
is a first step in connecting and understanding properties of GC systems, their host galaxies and their central supermassive black holes.

Surveys of elliptical galaxies show radial GC profiles to be less concentrated when compared to the galactic stellar light profiles
(\citealt{1979ARA&A..17..241H, 1996ApJ...467..126F, 1999AJ....117.2398M, 2009A&A...507..183C} and references therein).
An impressive example is the 10 kpc core in the spatial distribution of GCs in NGC~4874, one of the two
dominant elliptical galaxies inside the Coma cluster \citep{2011ApJ...730...23P}. Two competing scenarios attempt to explain these cored distributions.
In one scenario, the core originated from processes operating at the onset of galaxy formation
in the very early universe \citep{1986AJ.....91..822H, 1993ASPC...48..472H}. The other scenario assumes that GCs were formed co-evally with the field stars
with a cuspy distribution, i.e. comparable to the galactic stellar light profile. In this second scenario, the observed cores in the GC distributions
are caused by the subsequent erosion and destruction of globular clusters in the nucleus of the galaxy itself \citep{1993ApJ...415..616C, 1998A&A...330..480B,
2000MNRAS.318..841V, 2003ApJ...593..760V}. It is this scenario we would like to shed light on with this study. 

However, taking all the relevant processes that affect the GC erosion rates in elliptical galaxies into account is numerically
challenging. This is due to the fact that there are several internal and external processes acting simultaneously on the dissolution of globular
clusters, such as two-body relaxation, stellar mass loss and tidal shocks \citep{1997ApJ...474..223G, 1997MNRAS.289..898V, 2001ApJ...561..751F, 2003MNRAS.340..227B, 2006MNRAS.371..793G}.
In this study, we present a new code named \textsc{Muesli} to investigate several processes that dominate cluster erosion:
(i) tidal shocks on eccentric GC orbits and relaxation driven dissolution, and their dependence on the anisotropy profile of the GC population, 
(ii) tidal destruction of GCs due to a central super-massive black hole
(iii) stellar evolution and
(iv) orbital decay through dynamical friction. That is:

(i) GCs lose mass when stars get beyond the limiting Jacobi radius, $r_{J}$, and become unbound to the cluster. Two-body relaxation will
cause any GC to dissolve with time. The dissolution time depends on the mass and extent of the GC as well as the strength of the tidal field \citep{2003MNRAS.340..227B}. 
GCs on very eccentric orbits are particularly susceptible for disintegration within a few orbits owing to the strong tidal forces near
the galactic center. In radially biased velocity distributions, large fractions of orbits are occupied by such eccentric, i.e. low
angular momentum orbits, and the overall destruction rate of globular clusters is strongly enhanced over the isotropic case. The same holds
for triaxial galaxies when globular clusters move on box orbits \citep{1989MNRAS.241..849O, 1997MNRAS.292..808C, 2005MNRAS.356..899C}. 

(ii) The gradient of the potential which is relevant for the destruction of GCs is increased by the presence of supermassive
black holes. SMBHs are commonly found in the cores of luminous galaxies \citep{1998AJ....115.2285M,
2007ApJ...662..808L} and the connection between SMBHs and globular clusters is of particular interest. \citet{2010ApJ...720..516B}
and \citet{2011MNRAS.410.2347H} found empirical relations between the total number of GCs and the mass of the central black hole.
The origin of this linear $\mbh-N_{\scriptsize{\mbox{GC}}}$ relation is under debate. See \cite{2013arXiv1312.5187H} for a most recent version
of the $\mbh-N_{\scriptsize{\mbox{GC}}}$ relation and comparison to other globular cluster/host galaxy relations.
There is some evidence that $\mbh$ and $N_{\scriptsize{\mbox{GC}}}$ are indirectly coupled over the properties of their host galaxies
\citep{2012AJ....144..154R}, however a direct causal link cannot be ruled out owing to the difficulty of studying
the growth of SMBHs from accreted cluster debris.

(iii) Another effect is mass loss by stellar evolution (SEV). SEV decreases the globular cluster mass most significantly during an initial phase of
roughly $100$ Myr. In this period, O and B stars lose most of their mass through stellar winds and supernovae (e.g.~\citealt{2008sse..book.....D}). Over a Hubble time, a
stellar population loses about 30-40\% of its mass due to stellar evolution \citep{2003MNRAS.340..227B}

(iv) Finally, massive objects like globular clusters lose energy and angular momentum due to dynamical friction (DF) when migrating
through an entity of background particles. GCs will gradually approach the center of the galaxy where they are destroyed efficiently
as described above. In low luminosity spheroids ($L\approx10^{10}\lsun$) decaying GCs might also merge together and contribute to
the growth of nuclear star clusters \citep{1975ApJ...196..407T, 2011ApJ...729...35A, 2013ApJ...763...62A, 2013arXiv1308.0021G}. Among other quantities,
the efficiency of DF depends on the departure of the host galaxy from spherical symmetry \citep{2004MNRAS.349..747P}, and becomes
largest for low angular momentum orbits \citep{1992MNRAS.254..466P, 2005MNRAS.356..899C}.

Like a real muesli, our \underline{Mu}lti-Purpose \underline{E}lliptical Galaxy \underline{S}CF + Time-Transformed \underline{L}eapfrog
\underline{I}ntegrator (\textsc{Muesli}) consists of several well chosen ingredients. \textsc{Muesli} has a high flexibility and is designed for
computing GC orbits and erosion rates in live galaxies.
It can handle spherical, axisymmetric and triaxial galaxies with arbitrary density profiles, velocity distributions
and central SMBH masses for which no analytical distribution functions exist. Since the potential of the galaxy is computed
self-consistently, the code can handle time evolving potentials due to e.g. the interaction of the galaxy and a central black hole \citep{1998ApJ...498..625M}
or even non-virialised structures.

\textsc{Muesli} is designed to constrain the field-star and GC formation efficiencies in the early universe. This can be done by relating
the computational outcomes with observations of the GC specific frequency, $S_{N}$, which is the number of observed globular
clusters normalized to to total mass/luminosity of the host galaxy \citep{2010MNRAS.406.1967G, 2013ApJ...772...82H, 2013arXiv1307.6563W}. The U-shaped $S_{N}$
distribution, being highest for the least massive and most massive galaxies, traces the impact of feedback processes operating in
different galactic environments. However, the quantitative examination of these processes requires knowledge about the total fraction
of GCs eroded over time.

In this first paper, we provide detailed information about the code and about $N$-body model generation, and we show results from
the code testing. We apply our code to erosion processes of GCs inside spherical galaxies with Hernquist and S\'{e}rsic profiles, isotropic and
radially biased velocity distributions and central SMBHs. This is done for four representative galaxies. These galaxies cover a wide range
of masses ($M_{\scriptsize{\mbox{GAL}}} \approx\unit{10^{9}-10^{12}}{\msun}$), sizes ($R_{e}\approx\unit{10^{2}-10^{4}}{\mathrm{pc}}$) and central
SMBH masses ($\mbh\approx \unit{10^{6}-10^{10}}{\msun}$). Erosion rates in axisymmetric and triaxial galaxies, as well as
nuclear star cluster and SMBH growth processes by cluster debris are reserved for later publications. 

The present paper is organized as follows. The \textsc{Muesli} code and the dynamics governing globular cluster dissolution and disruption
processes are specified in \S~\ref{sec:method}. At the end of this section we introduce the initial conditions of the GCs
and discuss the generation of the underlying galaxy models. Results are presented in \S~\ref{sec:results}, followed by a
critical discussion (\S~\ref{sec:critical_discussion}). The main findings are summarized in \S~\ref{sec:conclusion}. Extensive tests
of the code are carried out in the Appendix \S~\ref{sec:testing}.

\section{METHOD}\label{sec:method}
In the following we briefly describe the main ingredients of \textsc{Muesli}. These ingredients can easily
be modified, exchanged or upgraded, making \textsc{Muesli} a versatile platform for the study of GC dynamics in elliptical galaxies and related problems.
\subsection{SCF Integration Method and Scaling Issues}\label{subsec:software&scaling_scf}
The computations are performed with the self-consistent field (SCF) method \citep{Hernquist1992}. The SCF algorithm
uses a basis-function approach to evaluate an expression for the potential $\phi$ from the underlying matter configuration. The
orders $n, l$ of the radial and angular expansion terms can be adjusted to match the type of galaxy.
In the underlying study we restrict ourselves to spherically symmetric galaxies.
The usage of $l>0$ in spherical galaxy models can result in inhomogeneities and unphysical drifts of the angular momentum vectors.
Therefore $n=30, l=0$ is adopted for the main computation of the spherical galaxies of this study, while $n=l\geq10$ is chosen for axisymmetric
and triaxial galaxies. Tests are performed in \S~\ref{subsec:testing_discreteness} and \S~\ref{subsec:testing_phasespace}. 

The particle trajectories are integrated forward in time with the time-transformed leapfrog (TTL) scheme \citep{2002CeMDA..84..343M}
combined with an iteration method to account for the inclusion of external (velocity-dependent) forces. These forces are the dynamical
friction force (\S~\ref{subsubsec:df}) and post Newtonian forces allowing to mimic general relativistic effects arising from the central SMBH. 
Post-Newtonian terms are not relevant for GC destruction processes as they occur on distances much larger than event horizon scales where
GR effects are negligible and so we do not consider them for this study.

While the code uses conventional model units, $M_{\scriptsize{\mbox{GAL}}}=R_{H}=G=1$, the relevant globular cluster quantities
(\S~\ref{subsec:init_cond_gc_mass_size}) as well as the dynamical friction force (\S~\ref{subsubsec:df}) are defined in physical
dimensions. The scaling of time, mass and size is performed during computations.

\subsection{GC Dissolution Mechanisms}\label{subsec:GC_dissolution_mechanisms}
We aim at quantifying the relevance of internal and external effects like stellar evolution (\S~\ref{subsubsec:dyn_of_GSs_relaxation}),
relaxation driven evolution of GCs in tidal fields (\S~\ref{subsubsec:dyn_of_GSs_relaxation}), tidal disruption through shocks (\S~\ref{subsubsec:tidal_shocks})
and dynamical friction (\S~\ref{subsubsec:df}) for shaping a cored GC distribution.

\subsubsection{Stellar Evolution and Two-Body Relaxation}\label{subsubsec:dyn_of_GSs_relaxation}
Stellar evolution reduces the cluster mass most significantly during an initial phase of
roughly $100$ Myr. During this period, the most massive stars lose mass through stellar winds and supernovae (e.g.~\citealt{2008sse..book.....D}). 
 
We implemented the combined scheme for stellar evolution and energy-equipartition driven evaporation in tidal fields from \cite{2003MNRAS.340..227B}.

Compared to the long-term dynamical evolution, SEV decreases the initial cluster mass nearly instantaneously
by $30\%$ (\citealt{2003MNRAS.340..227B}, their Figure 1). Therefore, initial SEV can be taken into account by using a constant
GC mass correction factor of 0.70 (\citealt{2003MNRAS.340..227B}, their Equation 12). 

On the other hand, two-body relaxation causes a more continuous mass loss \citep{1961AnAp...24..369H, 2003MNRAS.340..227B, 2003gmbp.book.....H}.
Due to two-body encounters, stars gain enough energy so that they can leave the cluster. In the long term, this process leads to the dissolution
of any star cluster. The process of relaxation-driven mass loss is accelerated when globular clusters are embedded in the external
tidal field of a host galaxy as the potential barrier for escape is lowered \citep{1990ApJ...351..121C,
2003MNRAS.340..227B}. In this case, a star may separate from the GC when passing beyond a characteristic radius,
commonly known as the Jacobi radius, $r_{J}$ (e.g.~\citealt{1962AJ.....67..471K, 1987degc.book.....S}).
When the cluster moves on a non-circular orbit the tidal field strength varies with time, and so does the Jacobi radius.
With growing eccentricity of the cluster's orbit, the extrema between the Jacobi radius at
perigalacticon, where it is smallest, and apogalacticon increase (see e.g.~\citealt{2010MNRAS.407.2241K,
2013ApJ...764..124W}).

\cite{2003MNRAS.340..227B} suggests that the time dependent mass loss rate of GCs in tidal fields can be approximated by a
linear function. A few modifications were added to account for arbitrary galactic density profiles and changing GC orbits
caused by dynamical friction. This is done by first computing the galactocentric distance, $r_{G}$, and velocity, $v_{G}=\sqrt{a\cdot r_{G}}$, where $a=\left|\vec{a}\left(\vec{r}\right)\right|$
is the acceleration at the position of the globular cluster. Then, the dissolution time is calculated using Eq.~7 of \citet{2003MNRAS.340..227B},
\begin{equation}\label{dissolutiontime}
 \frac{t_{\scriptsize{\mbox{DISS}}}}{\mbox{\small{[Myr]}}}=\beta\left(\frac{N_{0}}{\ln(0.02N_{0})}\right)^{\gamma}\frac{r_{G}}{\mbox{\small{[kpc]}}}
 \left(\frac{v_{G}}{220\mbox{\small{km/s}}}\right)^{-1},
\end{equation}
for every cluster individually. The two parameters $\beta$ and $\gamma$ depend on the concentration of the globular clusters. We chose values of $\beta=1.91$
and $\gamma = 0.75$ for our main computations (as well as $\beta=1.21$ and $\gamma = 0.79$ for a particular model), which have been found to reproduce the mass evolution of clusters with
a King density profile and a $W_0$ of 5.0. This is typical density profile among Milky-Way globular clusters (e.g.~\citealt{2005ApJS..161..304M}).
Depending on the density profile of the respective globular cluster, $\beta$ and $\gamma$ can change.
The initial number of globular cluster stars, $N_{0}$, is approximated by $N_{0}=m_{\scriptsize{\mbox{GC}}}\left(t=0\right)/0.547$ \citep{2003MNRAS.340..227B}.
We assume that mass is lost linearly with time, so after each timestep, $\Delta t$, the GC
mass is reduced by an amount $\Delta m=\Delta t\cdot m_{\scriptsize{\mbox{GC}}}\left(t=0\right)/t_{\scriptsize{\mbox{DISS}}}$. The galactocentric distance used in Eq.~\ref{dissolutiontime}
gets updated at each peri- or apocenter passage and we use the last pericenter or apocenter distance for $r_{G}$. In this way our method reproduces the ($1-\epsilon$) scaling
of the lifetimes found by \cite{2003MNRAS.340..227B} without having to calculate orbital parameters.

In galaxies where the circular velocity $v_{G}$ varies with radius, we use an average of the pericenter and apocenter velocity $\overline{v}=\left(v_{r_{apo}}+v_{r_{peri}}\right)/2$
to account for the varying circular velocity. Once the mass of a cluster is less than $m_{\scriptsize{\mbox{GC}}}=\unit{10^{2}}{\msun}$, it is assumed to be dissolved.

\subsubsection{Tidal Shocks}\label{subsubsec:tidal_shocks}
The variation of the tidal field does not only enhance the overall mass
loss rate but also increases the cluster's internal energy \citep{1997ApJ...474..223G, 1999ApJ...514..109G}.
If the pericentre distance to the galactic center is small, the energy input through tidal variation acts like a
tidal shock and the energy gain of the cluster is significant \citep{1999ApJ...513..626G, 2010MNRAS.401..105K,
2013ApJ...764..124W, 2013MNRAS.tmp.1577S}. In these cases mass loss of the cluster gets driven by 
tidal shocks leading to quick cluster dissolution \citep{1997MNRAS.289..898V, 1997ApJ...474..223G,
1999ApJ...514..109G, 2009MNRAS.399.1275P, 2010MNRAS.407.2241K}. We found that by using a second disruption criterion in addition to
\S~\ref{subsubsec:dyn_of_GSs_relaxation}, we can compensate for the underestimated mass-loss rate for clusters
on very eccentric orbits within stong tidal fields and obtain better fits to direct \textsc{Nbody6} integrations (Fig.~\ref{massevolu}). This criterion is derived below: 

\begin{figure}
\centering
\includegraphics[width=8.5cm]{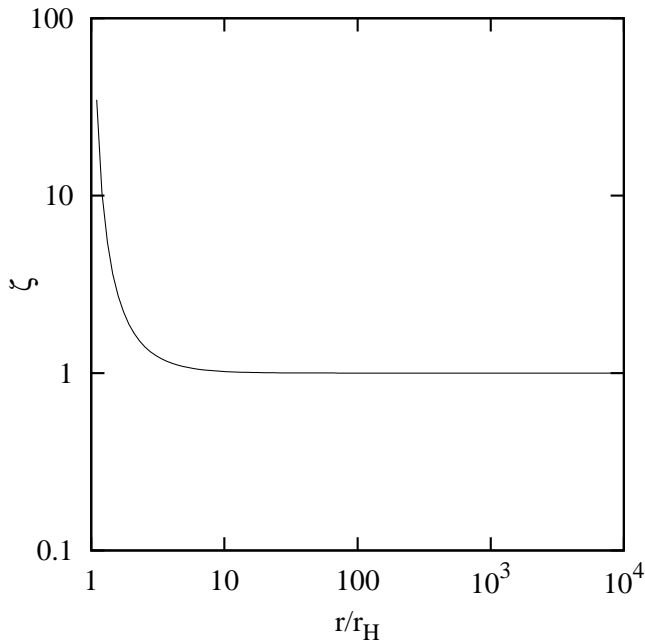}
\captionsetup{format=plain,labelsep=period,font={small}}
\caption{The ratio $\zeta=\frac{a(r+r_{H})-a(r-r_{H})}{2r_{H}}\Big/\left(\frac{\partial^{2} \phi}{\partial
r^{2}}\right)^{-1}$ plotted for a $\phi\propto -\frac{1}{r}$ potential. The
distance $r$ is given in GC half-mass radii, $r_{H}$. The
approximation $\frac{\partial^{2} \phi}{\partial r^{2}}\approx \frac{a(r+r_{H})-a(r-r_{H})}{2r_{H}}$ is accurate
down to a few $r_{H}$ and is almost exact for mass configurations with less steeply varying gradients.}
\label{effpot1}
\end{figure}
For quantifying the strength of tidal shocks we
calculate the Jacobi radius, $r_{J}$, following \citet{1962AJ.....67..471K}:
\begin{equation}\label{f1}
 r_{J}= \left(\frac{Gm_{\scriptsize{\mbox{GC}}}}{\Omega^{2}-\frac{\partial^{2} \phi}{\partial r^{2}}}\right)^{\frac{1}{3}}.
\end{equation}
Here $\Omega=\frac{\left|\vec{r}\times \vec{v} \right|}{r^{2}}$ is the cluster's orbital galactocentric
angular velocity. We approximate the second spatial derivative of the galactic potential in Eq.~\ref{f1} by 
\begin{equation}\label{diff_qout}
\frac{\partial^{2} \phi}{\partial r^{2}}\approx \frac{a(r+dr)-a(r-dr)}{2dr},
\end{equation}
where $a(r)$ is the acceleration at galactocentric radius $r$, and $dr<<r$ is a sufficiently small distance
away from the cluster center. We choose $dr=r_{H}$ to be the 3D half-mass radius\footnote{For galaxy specific quantities like the half
mass radius, $R_{H}$, we use capital letters.} of the respective globular cluster. This approximation works well even for the
quickly declining $1/r$ potential close to the central SMBH (see Figure~\ref{effpot1}). If a GCs falls below the galactic center distance $r=2r_{H}$,
$r_{J}$ is taken to be the $r_{J}$ at $r=2r_{H}$. 

Equation~\ref{f1} is valid for star clusters on eccentric orbits in arbitrary spherical potentials
(see also, e.g., \citealt{1987degc.book.....S, 2006MNRAS.366..429R, 2009MNRAS.392..969J, 2010MNRAS.401..105K,
2011MNRAS.418..759R, 2013MNRAS.429.2953E}). It has been compared to $N$-body simulations of dissolving star clusters
and found to well reproduce the radius at which stars escape from the star cluster into the tidal tails \citep{2012MNRAS.420.2700K}.

\begin{figure}
\centering
\includegraphics[width=8.5cm]{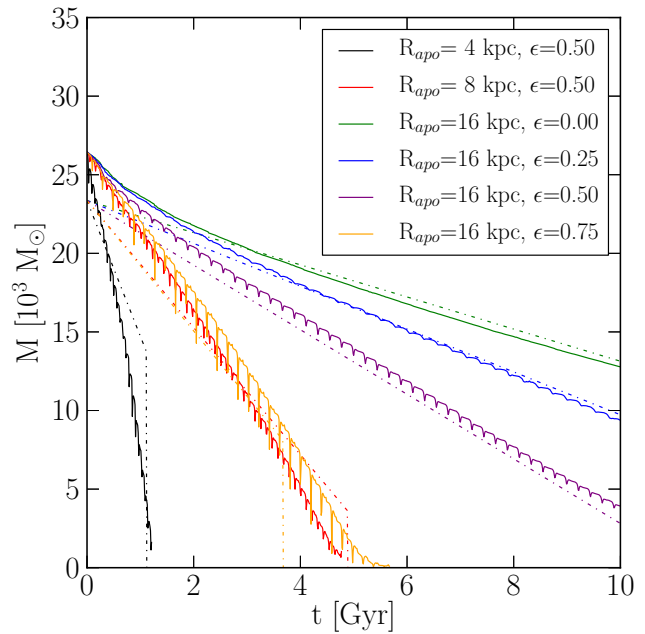}
\captionsetup{format=plain,labelsep=period,font={small}}
\caption{Temporal evolution of the cluster mass. The data were taken from a set of direct $N$-body
experiments of a number of star clusters on orbits with different eccentricities in the tidal field of a galaxy.
Clusters evolve from their initial masses of $\approx 3\times\unit{10^{4}}{\msun}$ towards dissolution. The clusters which
are affected by the strongest tidal shocks (solid black, orange and red lines) are subject to fast dissolution even though their
initial values $x=r_{H0}/r_{J}$ started below $x=0.5$. The theoretically predicted mass evolutions as described in
\S~\ref{subsubsec:dyn_of_GSs_relaxation} and \S~\ref{subsubsec:tidal_shocks} are shown with dashed lines.}
\label{massevolu}
\end{figure}
\begin{figure}
\centering
\includegraphics[width=8.5cm]{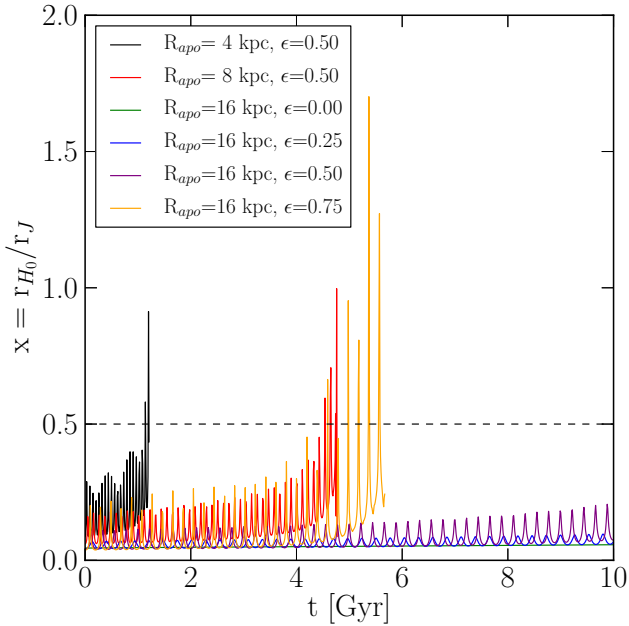} 
\captionsetup{format=plain,labelsep=period,font={small}}
\caption{Ratio of initial half-mass radius $r_{H0}=\,4\mbox{pc}$ over Jacobi radius $r_{J}$ versus time $t$ for the models shown in Fig.~\ref{massevolu}. 
Depending on their orbits, the ratio $x$ may temporarily reach values of $x>0.5$, but those clusters are
subject to quick dissolution. Surviving clusters stay well below our disruption threshold of $x=0.5$
(dashed line) for the simulation length of 10 billion years.}
\label{RhRj}
\end{figure}
\begin{figure}
\centering
\includegraphics[width=8.5cm]{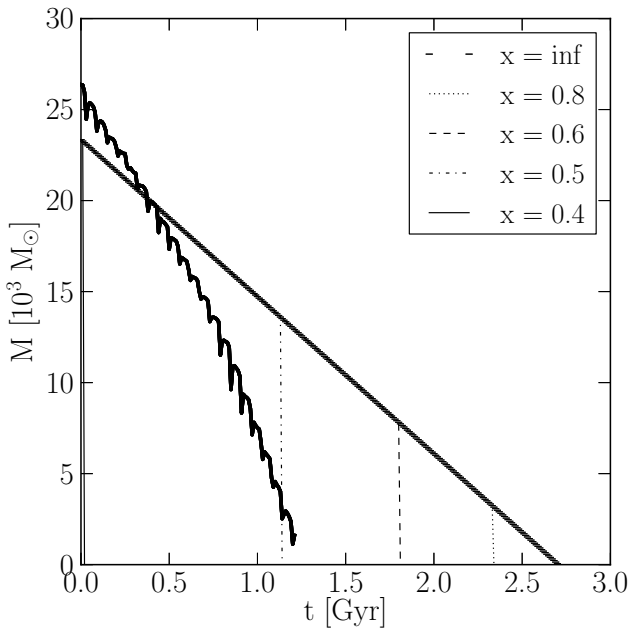}
\captionsetup{format=plain,labelsep=period,font={small}}
\caption{Mass evolution of a cluster with an apogalactic distance of 4 kpc and an orbital eccentricity of 0.5 (thick solid line).
Also shown are the theoretical mass evolutions for different values of $x$. If $x$ is chosen too small, the disruption by tidal forces is overestimated,
if it is chosen too large, the cluster survives for too long.}
\label{x_criterion}
\end{figure}
We assume that a cluster is disrupted if its ratio $x=\frac{r_{H}}{r_{J}}$ is larger than $x=0.5$ at any point during
its orbit\footnote{Furthermore we assume that GCs passing a SMBH within their 3D half mass radii, $r_{H}$, are disrupted as well.}.
This approach yields a safe lower limit on the disruption rate as some cluster would be confronted with
tidal fields even in excess of $x=0.5$ at perigalacticon. 
The motivation behind using the limit $x=0.5$ is subject to (i) observations of GCs
in the Milky Way and (ii) direct $N$-body computations. The majority of GCs in the Milky Way are on eccentric orbits
\citep{1999AJ....117.1792D}. Most of them have ratios, $x$, well below 0.2, while only one GC has $x>0.5$ \citep{2010MNRAS.401.1832B,
2013MNRAS.429.2953E}. The one cluster with $x\approx 0.55$ is the low-mass globular cluster Pal 5, which is thought
to be in the very final stages of dissolution due to its pronounced tidal tails \citep{2003AJ....126.2385O,
2004AJ....127.2753D}. Observations therefore suggest a limit of $x=0.5$ to be reasonable. In addition to that we also
performed direct $N$-body computations with the \textsc{Nbody6} code \citep{Aarseth1999, Aarseth2003} on the GPU
computers of the SPODYR group at the AIfA, Bonn. We ran 32 simulations of compact and massive star clusters
($m_{\scriptsize{\mbox{GC}}}>10^4\msun$) on a range of orbits within a galactic tidal field. We followed their dynamical evolution 
for 10 billion years or until total dissolution, skipping the first 1 Gyr in which the clusters' evolution is dominated by the SEV
processes and expansion as a consequence of rapid mass loss. See Fig.~\ref{massevolu} for a representative sample.
Also shown in Fig.~\ref{massevolu} is the theoretically predicted mass evolution using Eq.~\ref{dissolutiontime} and the disruption criterion $x=0.5$. Our model
clusters lie in between a King profile with $W_0 = 5.0$ and $W_0 = 3.0$ so we had to use $\beta = 1.21$ and $\gamma = 0.79$ for this
comparison\footnote{To quantify the dependency of the overall GC system erosion rate on internal cluster profiles,
computations with $\beta = 1.91$ and $\gamma = 0.79$ (i.e. $W_0 = 5.0$) and $\beta = 1.21$ and $\gamma = 0.79$ (green line in Figure~\ref{Erosionplot})
were performed.}.

As can be seen in Fig.~\ref{RhRj}, some of the clusters
evolve from initial ratios $x=\frac{r_{H0}}{r_{J}}=\frac{4\scriptsize{\mbox{pc}}}{r_{J}}\approx0.3$ quickly to $x=0.5$ where they are destroyed very rapidly.
Only clusters with ratios well below $x=0.5$ have a chance to survive for more than a Hubble time.
Similar results have also been found by \citet{2007MNRAS.374..344T} and \citet{2008MNRAS.389..889K}. Hence, $N$-body computations
also suggest that a value of $x_{\tiny{\mbox{crit}}}=0.5$ is a conservative limit for our computations. This criterion shows in a
clear manner which areas in the phase space cannot be stably populated by clusters of a given mass.
Even more so, because we neglect in our treatment of disruption processes that the half-mass radius grows with time when
the clusters are initially tidally underfilling \citep{2010MNRAS.408L..16G, 2012ApJ...756..167M, 2013ApJ...764..124W}.
In addition to that, it should be noted that the present study investigates GC dissolution and disruption processes in spherical
galaxies. Here the angular momentum of cluster orbits, apart from dynamical friction, is a conserved quantity. A single GC not being
fully disrupted once the tidal field strength exceeds $x=0.5$ would be destroyed within the next few orbits.

As can be seen in Fig.~\ref{x_criterion}, increasing $x$ allows clusters on very eccentric orbits and deep within the tidal field
of their host galaxy to survive longer than for $x=0.5$ and longer than found in direct $N$-body simulations. Thus, we may then overestimate the
number of surviving clusters in the central part of the host galaxies. A value smaller than 0.5 on the other hand may be too weak.
Therefore, computations with more restrict criteria ($x=0.8$ and $x=\infty$ i.e. no tidal disruption) but otherwise identical physical
properties are performed as well. Our additional simulations allow us to constrain the systematics introduced by our second disruption criterium.

\subsubsection{Dynamical Friction}\label{subsubsec:df}
Massive objects moving through a background of particles will decelerate and lose orbital energy by dynamical
friction (DF) \citep{1943ApJ....97..255C}. This effect may have profound implications for the orbital evolution and,
hence, the fate of globular clusters. Our \textsc{Muesli} code is designed to obtain the impact of DF on the destruction of GCs in
galaxies with isotropic and anisotropic velocity distributions as well as axisymmetric and triaxial galaxies. 

The equation of motion of a massive object like a globular cluster in a galaxy with DF, is:
\begin{equation}\label{f2}
 \vec{a}_{\scriptsize{\mbox{GC}}}\left(\vec{r}\right)=-\vec{\nabla}\phi(\vec{r})+\vec{a}_{\scriptsize{\mbox{GC}},\
 \scriptsize{\mbox{DF}}}\left(\vec{r}\right).
\end{equation}
Here, $\vec{a}_{\scriptsize{\mbox{GC}}}$ is the total acceleration of the globular cluster, $-\vec{\nabla}\phi(\vec{r})$ is
the acceleration due to the combined galactic and SMBH potential, while $\vec{a}_{\scriptsize{\mbox{GC}},\ \scriptsize{\mbox{DF}}}$
describes the deacceleration due to DF. Chandrasekhar's dynamical friction formula has been extended to account for ellipsoidal velocity distributions by
\cite{1992MNRAS.254..466P} and $\vec{a}_{\scriptsize{\mbox{GC}},\
\scriptsize{\mbox{DF}}}\left(\vec{r}\right)$ has the form:
\begin{align}\label{f3}
 \vec{a}_{\scriptsize{\mbox{GC}},\ \scriptsize{\mbox{DF}}}\left(\vec{r}\right)=&-\gamma_{1}\left(\vec{r}\right)\tilde{v}_{1}\left(\vec{r}\right)\vec{e}_{1}-
  \gamma_{2}\left(\vec{r}\right)\tilde{v}_{2}\left(\vec{r}\right)\vec{e}_{2} \\
   & -\gamma_{3}\left(\vec{r}\right)\tilde{v}_{3}\left(\vec{r}\right)\vec{e}_{3}\notag\ , 
\end{align}
where the dynamical friction coefficients $\gamma_{i}\left(\vec{r}\right)$ can be written as:
\begin{align}\label{f4}
&\gamma_{i}(\vec{r})=\frac{2\sqrt{2\pi}\rho\left(\vec{r}\right)G^{2}m_{\scriptsize{\mbox{GC}}}\ln{\Lambda}}
 {\sigma^{3}_{1}\left(r\right)}\times \\
                    &\int_{0}^{\infty}\frac{\exp\left(-\frac{\tilde{v}_{1}^{2}\left(
                    \vec{r} \right)/\left(2\sigma^{2}_{1}\left(r\right)\right)}{1+u}
                    -\frac{\tilde{v}_{2}^{2}\left(\vec{r} \right)/\left(2\sigma^{2}_{2}
                     \left(r\right)\right)}{\epsilon^{2}_{2}+u}-\frac{\tilde{v}_{3}^{2}
                     \left(\vec{r} \right)/\left(2\sigma^{2}_{3}\left(r\right)\right)}{\epsilon^{2}_{3}+u} 
                     \right)}{\left(\epsilon^{2}_{i}+u  \right)\sqrt{\left(1+u \right) 
                     \left( \epsilon^{2}_{2}+u\right)\left(\epsilon^{2}_{3}+u \right)}}du\notag.  
\end{align}
The function $\Lambda$ that appears in the Coulomb logarithm, $\ln{\Lambda}$, can be obtained for bodies with a finite
size (\citealt{2008gady.book.....B}, their Eq. 8.2) by:  
\begin{equation}\label{f4b}
 \Lambda=\frac{b_{\scriptsize{\mbox{max}}}}{\mbox{max}\left( r_{H}, Gm_{\scriptsize{\mbox{GC}}}/v_{\scriptsize{\mbox{typ}}}^{2} \right)}. 
\end{equation}
\begin{figure}
\centering
\includegraphics[width=8cm]{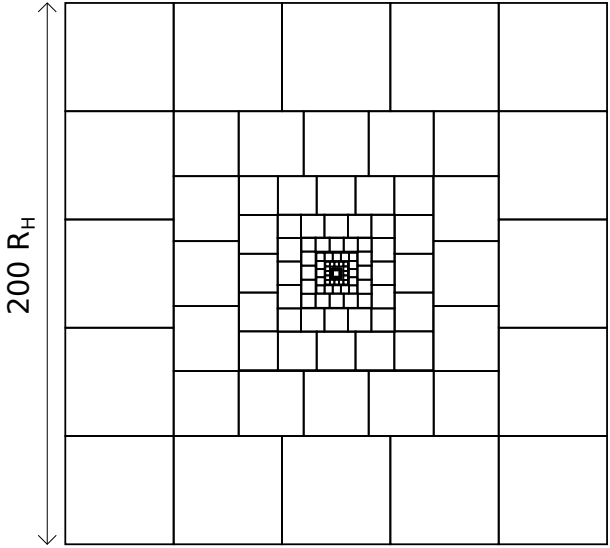}
\captionsetup{format=plain,labelsep=period,font={small}}
\caption{A 2D illustration of a 5x5x5 grid. The eigenvectors of the velocity dispersion tensor are calculated in
each cell. Grid cells are logarithmically refined towards the inner part of a galaxy to guarantee sufficient resolution. The size 
of the outer grid is chosen to be 200 times the size of the galactic half mass radius $R_{H}$ to encompass the whole galaxy.}
\label{cube}
\end{figure}

The maximum impact parameter $b_{\scriptsize{\mbox{max}}}$ is approximated by the galactocentric distance
$r$. This approach yields a more realistic treatment than by assigning a constant value for the Coulomb logarithm
\citep{2003ApJ...582..196H, 2003MNRAS.344...22S}. The validity of Eq.~\ref{f4b} is restricted to
$\Lambda>1$ in order to prevent unphysical acceleration by DF. The characteristic velocity is
$v_{\scriptsize{\mbox{typ}}}^{2}\approx GM_{\scriptsize{\mbox{GAL}}}/R_{\scriptsize{\mbox{H}}}$. Here $M_{\scriptsize{\mbox{GAL}}}$
and $R_{H}$ correspond to the total mass and half mass radius of the galaxy.
The parameter $\epsilon_{i}^{2}$ which appears in Eq.~\ref{f4} is given by the ratio of the eigenvalues $\sigma_{i}^{2}\left(\vec{r}\right)/\sigma_{1}^{2}\left(\vec{r}\right)$
of the velocity dispersion tensor $\sigma^{2}_{ij}= \overline{v_{i}v_{j}}-\overline{v}_{i}\overline{v}_{j}$.
For convenience, the velocity dispersion component with the largest eigenvalue of $\sigma^{2}_{ij}$ is defined to be $\sigma^{2}_{1}\left(r\right)$.
The integral is evaluated numerically for each integration timestep by using the Gau\ss-Legendre integration method in combination with
logarithmic mapping. The density $\rho\left(\vec{r}\right)$ is obtained directly from the SCF algorithm.
The velocity components $\tilde{v}_{i}\left(\vec{r}\right)=\cos{\theta_{i}}\left|\vec{v}_{\scriptsize{\mbox{GC}}}\left(
\vec{r}\right)\right|$ with $\cos{\theta_{i}}=\frac{ \vec{e}_{i}\cdot \vec{v}_{\scriptsize{\mbox{GC}}}\left(\vec{r}\right)}{\left|
\vec{e}_{i} \right| \cdot \left|\vec{v}_{\scriptsize{\mbox{GC}}}\left(\vec{r}\right) \right|}$ are obtained by the projection of the
GC velocity vector $\vec{v}_{\scriptsize{\mbox{GC}}}$ onto the normalized eigenvectors $\vec{e}_{i}$ of the velocity dispersion tensor $\sigma^{2}_{ij}$.
The position dependent eigenvalues $\sigma_{i}$ and eigenvectors $\vec{e}_{i}$ are calculated in
hundreds of cubic segments which are part of a 5x5x5 mesh with logarithmically increasing resolution towards the center (see Fig.~\ref{cube} for illustration).
This is achieved by replacing the inner 27 out of $125$ cubes by a second 5x5x5 grid. The procedure is repeated $G_{\scriptsize{\mbox{depth}}}\in \mathbb{N}$
times. The innermost resolution scale is $R_{\scriptsize{\mbox{res}}}=0.2R_{\scriptsize{\mbox{max}}}0.6^{G_{\scriptsize{\mbox{depth}}}}$.
The size $R_{\scriptsize{\mbox{max}}}$ of the outermost grid is chosen to encompass the whole galaxy. In this way a variable DF force acting on GCs in an elliptical galaxy
is handled. The underlying galaxy models are specified in Section~\ref{subsubsec:init_cond_gc_phasespace}.

Grid based calculations are always affected by discontinuities/jumps in combination with discreteness noise subject
to finite number of cells and particles. In order to counterbalance these systematics we apply the inverse distance weighting
(IDW) method \citep{Shepard:1968:TIF:800186.810616}. Irregularities are smoothed out by first calculating the center of mass of the particles in a box (which
is used as the position of the box), local eigenvalues and eigenvectors of $\sigma^{2}_{ij}$. Boxes containing only few particles are left out of
consideration. For the spatial interpolation only cells within a radius corresponding to the galactocentric distance
of an orbiting GC are taken into account. To guarantee that the contribution of the nearest box dominates, the weighting
parameter $p$ (also known as the power parameter) is calibrated in many $N$-body experiments to be $p=64$. For testing
issues we refer to \S~\ref{subsec:testing_grid}. The eigenvalues and eigenvectors are calculated at the beginning
of the computations and for each timescale the potential becomes updated by the SCF algorithm.

\subsection{Initial Conditions}\label{subsec:init_cond}
\subsubsection{Globular Cluster Mass and Size Distribution}\label{subsec:init_cond_gc_mass_size}
The present-day GC mass spectrum, $\mbox{d}N/\mbox{d}m\propto m^{-\beta}$, can be characterized as a power-law distribution with different
exponents for characteristic mass scales \citep{1994PASP..106...47M}. Usually, it is well approximated by the exponents
$\beta=0.2$ below and $\beta=2$ above a threshold mass of $m_{\scriptsize{\mbox{TH}}}=\unit{\left(1-2\right)\cdot10^{5}}{\msun}$. This two-component
power-law distribution resembles a bell-shaped function when expressed in terms of the number of globular clusters,
$\mbox{d}N$, per constant logarithmic cluster-mass interval, $\mbox{d}\log_{10} m$. For the initial cluster mass function,
we are using the single power-law distribution,
\begin{equation}\label{f5}
\frac{\mathrm{d}N}{\mathrm{d}m} \propto m^{-2}.
\end{equation}
It finds support by observations of young, luminous clusters in starburst galaxies where the mass spectrum monotonically follows a (single) power-law profile
with slope $\beta\approx2$ \citep{1994A&AS..104..379B, 1999ApJ...527L..81Z}\footnote{Recent investigations \citep{2009A&A...494..539L} found that the initial
mass distribution is also compatible with a Schechter-type mass function with a particular turn-down mass in the high GC mass regime. However, for simplicity
we use a single power-paw mass function here, as the differences will be limited to the high-mass end where only relatively few clusters are found.}.

It is our aim to investigate whether dissolution of low mass clusters is responsible for turning a power-law mass function 
into a bell shaped mass function (see also \citealt{1998A&A...330..480B, 2001ApJ...561..751F, 2003ApJ...593..760V, 2008ApJ...679.1272M, 2010ApJ...712L.184E}) by cluster disruption processes,
relaxation driven mass loss in tidal fields and dynamical friction. Scenarios involving gas expulsion \citep{2002MNRAS.336.1188K, 2007MNRAS.377..352P, 2008MNRAS.384.1231B} are not considered
in our main computations (with the exception of one model) and will be added in later publications. The overall GC mass range is chosen to be
$m_{\scriptsize{\mbox{GC}}}=\unit{10^{4}-10^{7}}{\msun}$. Clusters below$\unit{10^{4}}{\msun}$ are not considered, because in galaxies with an age of several
Gyr, they would have lost most (if not all) of their initial mass by energy-equipartition driven evaporation \citep{2003MNRAS.340..227B, 2010MNRAS.409..305L}. 

Observations find no strong correlation between half-mass radius and mass for GCs which are less massive than $\unit{10^{6}}{\msun}$ \citep{2005ApJ...627..203H,
2008MNRAS.386..864D}. The median 3D half-mass radius of GCs in typical early-type galaxies centers around $\overline{r_{H}}=
\unit{4}{\mbox{pc}}$ \citep{2005ApJ...627..203H, 2005ApJ...634.1002J}\footnote{For the conversion $r_{e}=0.75r_{H}$
\citep{1987degc.book.....S} of the projected half light radius, $r_{e}$, to the 3D half-mass radius, $r_{H}$, the mass-to-light
ratio, $\mathnormal{\Upsilon}(r)$, is assumed to be constant.}. The situation changes when the clusters become more massive than
a particular mass scale which is of the order of $\unit{10^{6}}{\msun}$\citep{2008MNRAS.386..864D}. Hence, we assume that they follow a trend given by: 
\begin{equation}\label{f10}
 r_{H}= \left\{ \begin{array}{l@{\quad:\quad}l}
 \unit{4}{\mathrm{pc}}  &   m_{\scriptsize{\mbox{GC}}} \lesssim  1.0\cdot10^{6}\msun \\
 \unit{4\left(\frac{m_{\scriptsize{\mbox{GC}}}}{\unit{10^{6}}{\msun}} \right)^{0.6}}{\mathrm{pc}}
 & m_{\scriptsize{\mbox{GC}}} \gtrsim 1.0\cdot10^{6}\msun
 \end{array}  \right. \
\end{equation}
The influence of other primordial size relations for GC erosion processes is not considered in this paper.

\subsubsection{Spatial Distribution of GCs}\label{subsubsec:init_cond_gc_phasespace}
Having defined cluster masses and sizes, the GC space and velocity vectors have to be distributed within the galaxies by making five underlying
assumptions\footnote{The code allows for individual adjustment of these aspects.}: 
\begin{enumerate}
  \item The initial GC phase space distribution equals the one of the underlying galaxy model
  \item Initial GC masses and sizes do not depend on the distance to the galactic center
  \item Accumulation of GCs through mergers or subsequent formation in star-forming events is neglected.
  \item The overall dynamics of the host galaxy are not influenced by globular cluster evolution processes
  \item All galaxy models are virialised and remain in isolation
\end{enumerate}
For this study we created several realistic base models. We assume that the stars follow a Sersic model \citep{Sersic1968} with concentration
$n=4$ and constant mass-to-light ratio, $\mathnormal{\Upsilon}$. They were generated by the deprojection of 2D S\'{e}rsic profiles into 3D density profiles.
Afterwards the density, potential and distribution function was calculated on a logarithmically spaced grid configuration of size $r\in \left [10^{-4},10^{2}\right]$
in model units, $G=R_{H}=M_{\scriptsize{\mbox{GAL}}}=1$. The distribution function for an anisotropic Osipkov-Merritt velocity profile
\citep{1979SvAL....5...42O, 1985AJ.....90.1027M} was calculated by making use of Equation (4.78a)
from \cite{2008gady.book.....B}. Here the velocity anisotropy parameter has the form $\beta(r)=\left(1+R_{A}^2/r^{2}\right)^{-1}$
and $R_{A}$ is the anisotropy radius. The particle positions were distributed according to the density profile, while
the normalized cumulative distribution function was used to allocate particle velocities. It was evaluated by the transformation
of the double integral into a single integral according to the substitution described in \cite{1985AJ.....90.1027M}\footnote{Their Equation 11.}.
Central SMBHs of mass $\mbh$ were implemented by adding the term $\phi_{BH}=-\mbh/r$ to the potential of the underlying mass distribution.
Afterwards, all particles were inverted (and doubled) through the origin. In this way the center of mass and density center were located at the point
of origin and the model stays at rest during computations. We also created galaxies following Hernquist \citep{1990ApJ...356..359H} and Jaffe \citep{1983MNRAS.202..995J} models.
Scale factors $a=\left(1+\sqrt{2}\right)^{-1}$ for Hernquist and $a=1$ for Jaffe models were used in order to fix the half mass radius to one.
\begin{table*}
\begin{center}
\begin{tabular}{|c|c|c|c|c|c|c|}  
 \hline
Model& Galaxy Example & $M_{\scriptsize{\mbox{GAL}}}$[$\unit{10^{9}}{\msun}$] & $R_{e}$[pc] & $R_{H}$[pc]  & $\mbh$[$\unit{10^{9}}{\msun}$] & Ref.\\ 
\hline
MOD1& M~32  & 0.8 & 125 & 170 & 0.0025 & 1,2,3\\
MOD2& NGC~4494  & 100 & 3715 & 5000 & 0.065 & 4,5,6\\
MOD3& IC~1459  & 300 & 6000 & 8050& 2.6 & 7,8\\
MOD4& NGC~4889  & 2000 & 25000 & 34000& 20 & 9,10 \\
\hline
\end{tabular}
\captionsetup{format=plain,labelsep=period,font={small}}
\caption{Adopted parameters for the simulated galaxies. $R_{H}=1.35R_{e}$ is used to calculate the 3D half mass radius $R_{H}$ from the effective
radius $R_{e}$. \textbf{References:} (1) \citet{1998AJ....115.2285M}; (2) \citet{2005AJ....129..712R}; (3) \citet{2004AJ....127.2031K};
(4) \citet{2007ApJ...664..226L} for $R_{e}$ and $M_{\scriptsize{\mbox{GAL}}}$ by assuming $\mathnormal{\Upsilon}=3$;
(5) \citet{1994MNRAS.269..785B}; (6) \citet{2011Natur.480..215M} black hole mass from $\mbh-\sigma$ relation;
(7) \citet{2004ApJ...604L..89H}, (8) \citet{2002ApJ...578..787C}, (9) \citet{2012ApJ...756..179M}, (10) \citet{2011Natur.480..215M}} 
\label{tablen=1}
\end{center}
\end{table*}
Finally, we generated an additional triaxially shaped model (required for testing issues of the dynamical friction routine) with a central core
and outer S\'{e}rsic $n=4$ profile from cold collapse computations \citep{1967MNRAS.136..101L, 1978MNRAS.185..227A,
1982MNRAS.201..939V, 1984ApJ...281...13M, 1998ApJ...498..625M}. A spherical distribution with a $\rho\propto r^{-1.5}$ density
profile and virial ratio $2T/\left|W\right|=0$ was set up for $0<r<2$. It collapsed and settled down into a strongly triaxial
configuration with $T=\left(a^{2}-b^{2}\right)/\left(a^{2}-c^{2} \right)=0.53$ within its half mass radius. Here $T$
is the triaxiality parameter and $a,b,c$ are the three main axes of the ellipsoidal configuration. It was evolved forward in
time with the \textsc{Nbody6} \citep{Aarseth1999, Aarseth2003} code until virialisation. The density center was shifted to
the center of origin and the model was rescaled to $R_{H}=1$. 
Models generated from collapse simulations are isotropic in their centers and radially biased at large galactocentric distances.

In the scenario of hierarchical structure formation (but see also \citealt{2004PASA...21..175S}), where smaller structures merge to build
up larger objects such as elliptical galaxies \citep{1977egsp.conf..401T, 1978MNRAS.183..341W, 1993MNRAS.264..201K, 2002NewA....7..155S}, violent relaxation causes
the merger products to be centrally isotropic and radially biased at large radii \citep{1967MNRAS.136..101L}. Our models agree with these cosmological 
predictions.

Related to the fact that a systematic scan over the fundamental plane of elliptical galaxies is beyond the scope of this paper, we
scaled our models to four representative elliptical galaxies. These are M~32, NGC~4494, IC~1459 and NGC~4889. While M~32 is a compact dwarf galaxy which is gravitationally bound to M~31, NGC~4889
is the most massive and extended galaxy in our sample. It is a brightest cluster galaxy (BCG) and defines together with NGC~4874 the gravitational center of the Coma cluster.
The four galaxies were chosen because they cover the full mass range of elliptical galaxies from small compact dEs to giant BCGs. They lie (within $10\%-45\%$ scatter) on
the $R_{e}-M_{\scriptsize{\mbox{GAL}}}$ relation (\citealt{2008MNRAS.386..864D}, their Equation 4) for low redshift bright elliptical galaxies, bulges and very compact
dwarf elliptical galaxies. We note that our results concerning globular cluster erosion processes in M~32 like compact galaxies should not be extrapolated to much more
extended dwarf spheroidal galaxies with weaker tidal fields (see Fig.2 in \cite{2008MNRAS.386..864D} and \S~\ref{subsubsec:TDDP} in this paper).
Complementary to the galactic mass range, our representative galaxies host central SMBH with masses in the range of a few $\unit{10^{6}}{\msun}$ (MOD1, i.e M~32)
up to $\unit{10^{10}}{\msun}$ (MOD4, i.e NGC~4889). The number of observed globular clusters ranges from 0 (M~32, \citealt{2013ApJ...772...82H}) to about
11.000 GCs (NGC~4889, \citealt{2009AJ....137.3314H}). The physical properties of all four galaxy models are summarized in Table~\ref{tablen=1}. 

\section{RESULTS}\label{sec:results}
In this section the main results of our computations are presented. They are divided into three major parts. In \S~\ref{sec:results_erosion_rate} 
we discuss general aspects of the globular cluster erosion rate in various galaxies. We present evidence for a new phase in the evolution of globular cluster
systems. In the following section (\S~\ref{sec:results_core}) we discuss the formation of cores in globular cluster systems. Finally, in \S~\ref{sec:mean_mass} we
investigate the evolution of the cluster mass function. 

\subsection{Globular Cluster Erosion Rate}\label{sec:results_erosion_rate}
\begin{figure*}
\centering
\includegraphics[width=19cm]{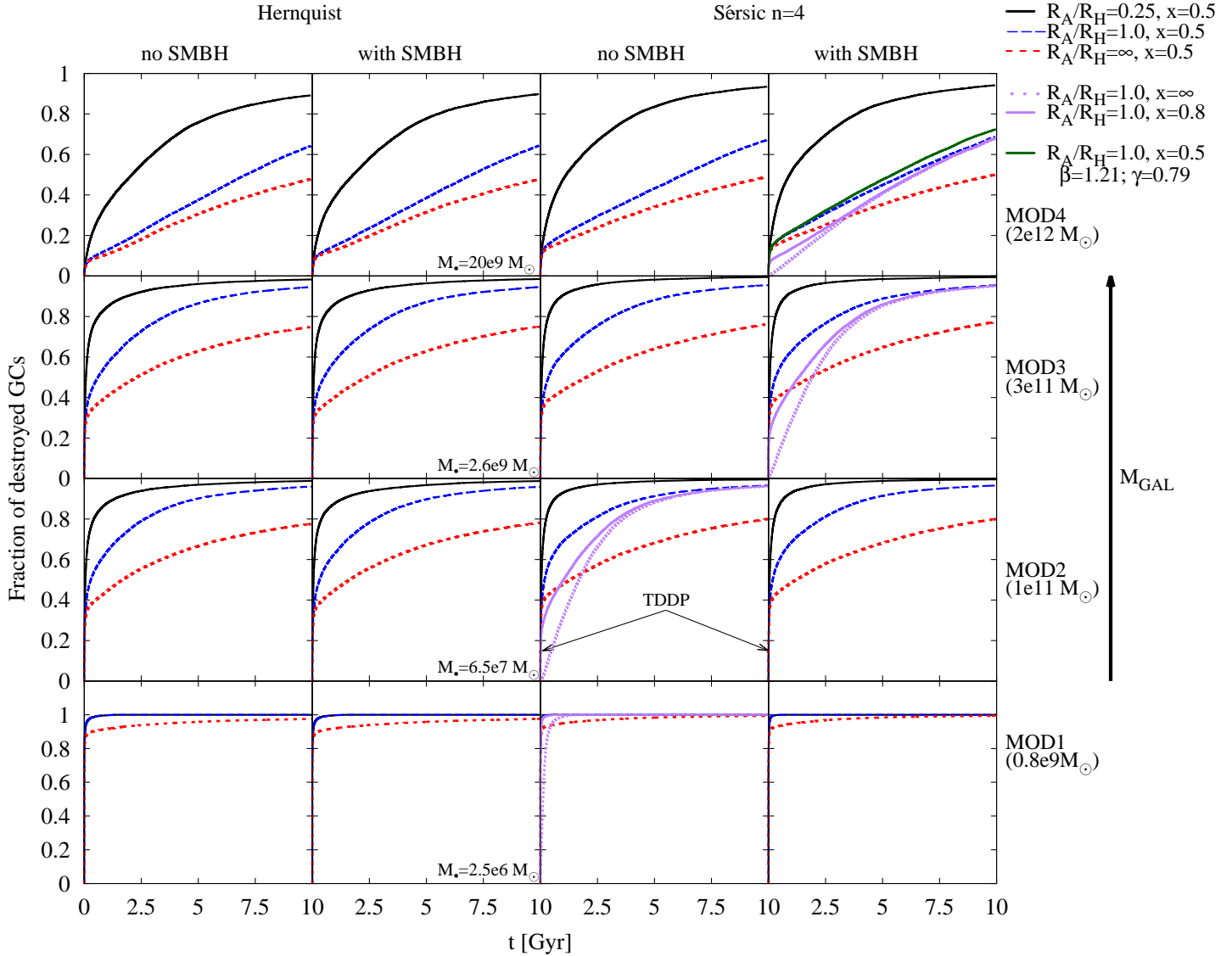}
\captionsetup{format=plain,labelsep=period,font={small}}
\caption{The temporal evolution of the fraction of destroyed globular clusters. Most of the destruction occurs at early times.  
Hence, we call this period the \textit{tidal disruption dominated phase (TDDP)}. It is followed by a relaxation-driven dissolution phase.
The destruction rate rises up to 100\% in models representing the least massive and compact galaxies (MOD1), independent of their velocity distribution.
The rate decreases to $50\%$ in the most luminous and extended galaxies (MOD4) with an isotropic velocity structure. The different
colors (i.e. black solid, blue long \& red short dashed lines) represent the degree of radial anisotropy characterized by the anisotropy radius $R_{A}$
of the Osipkov-Merritt models. Purple lines represent computations with more stringent criteria for GC destruction through
tidal shocks (\S~\ref{subsubsec:tidal_shocks}). Even by using $x=0.8$ or $x=\infty$ there is a \textit{TDDP} in the most compact models (MOD1).
With the exception of one model (green solid line, upper right panel) GCs were assumed to have King profiles with
a concentration parameter $W_0 = 5.0$. The green solid line illustrates the difference to a cluster population having parameters between $W_0=3.0$ and $W_0=5.0$, 
or more precisely by using $\beta=1.21$ and $\gamma=0.79$ in Equation~\ref{dissolutiontime} (\S~\ref{subsubsec:dyn_of_GSs_relaxation}).} 
\label{Erosionplot}
\end{figure*}
In order to evaluate the importance of the various processes for the dissolution of GCs, we performed 48 main computations plus additional 18 models
which are required to uncover more systematical effects. Each of these models consists of $2\cdot10^{7}$ stellar particles and 20,000 GCs, distributed according to a power-law mass
distribution as given by Eq.~\ref{f5} and the present-day half mass radius relation (Eq.~\ref{f10}). We chose to model 20.000 GCs in order to obtain
a good statistical significance of our results. The GCs in all computed models (with one exception) have King density profile with
concentration parameters $W_{0}=5$. This is a common quantity among globular clusters (\S~\ref{subsubsec:dyn_of_GSs_relaxation}). For this study we assumed that the
overall dynamics of the host galaxy are not influenced by globular cluster evolution processes
(\S~\ref{subsubsec:init_cond_gc_phasespace}) and hence all results can be scaled to different total GC numbers. For each of the four representative galaxies
shown in Table~\ref{tablen=1} we calculated models with and without central SMBHs, two different density profiles
(Hernquist and S\'{e}rsic $n=4$ models)\footnote{While at large radii both models agree well with each other, S\'{e}rsic $n=4$ models are centrally
more concentrated.} and three different velocity anisotropies ($R_{A}/R_{H}=0.25;1;\infty$), leading to 12 models per galaxy in total. 
We chose $R_{A}/R_{H}=0.25$ as the most radially biased model since below this limit galaxies become unstable due to a lack of tangential pressure
\citep{1985MNRAS.217..787M}. The models with $R_{A}/R_{H}=\infty$ represent the isotropic case. The central SMBHs masses were adopted from Table~\ref{tablen=1}.

We evolved all models for 10 Gyr under the influence of the generalized dynamical friction force (\S~\ref{subsubsec:df}).
As described in \S~\ref{subsubsec:tidal_shocks} \& ~\ref{subsubsec:dyn_of_GSs_relaxation}, clusters were assumed to be destroyed if: (i) the strength of the tidal
field, $x=\frac{r_{H}}{r_{J}}$, exceeded $x=0.5$, (ii) relaxation driven mass loss in tidal fields (and SEV) decreases their masses below the limit $m_{\scriptsize{\mbox{GC}}}=\unit{100}{\msun}$.
The temporal evolution of the globular cluster destruction rate is plotted in Fig.~\ref{Erosionplot} and absolute numbers of destroyed GCs after 10 Gyr evolution are summarized in Table~\ref{GCDIS}.
\begin{table*}
\begin{center}
\begin{tabular}{|c|c|c|c|c|c|c|} 
\hline
Model & Example Galaxy& Profile &$\mbh$[$\unit{10^{9}}{\msun}$] &$\chi_{0.25}$ & $\chi_{1}$ & $\chi_{\infty}$ \\ 
\hline
MOD4 & NGC~4889 & Hernquist & 0 &$0.89\pm0.01$ &$0.64\pm0.01$ &$0.48\pm0.01$\\
MOD4 & NGC~4889 & Hernquist & 20 &$0.90\pm0.01$ &$0.64\pm0.01$ &$0.48\pm0.01$\\
MOD4 & NGC~4889 & S\'{e}rsic n=4 & 0 &$0.93\pm0.01$ &$0.67\pm0.01$ &$0.49\pm0.01$\\
MOD4 & NGC~4889 & S\'{e}rsic n=4 & 20 &$0.94\pm0.01$ &$0.69\pm0.01$ &$0.50\pm0.01$\\

MOD3 &IC~1459& Hernquist & 0 &$0.98\pm0.01$&$ 0.95\pm0.01$ &$0.75\pm0.01$\\
MOD3 &IC~1459& Hernquist & 2.6 &$0.98\pm0.01$&$0.95\pm0.01$ &$0.75\pm0.01$\\
MOD3 &IC~1459& S\'{e}rsic n=4 & 0 &$0.99\pm0.01$&$0.95\pm0.01$ &$0.76\pm0.01$\\
MOD3 &IC~1459& S\'{e}rsic n=4 & 2.6 &$0.99\pm0.01$&$0.95\pm0.01$ &$0.77\pm0.01$\\

MOD2 &NGC~4494& Hernquist & 0 &$0.99\pm0.01$&$0.96\pm0.01$ &$0.78\pm0.01$\\ 
MOD2 &NGC~4494& Hernquist & 0.065 &$0.99\pm0.01$&$0.96\pm0.01$&$0.78\pm0.01$\\
MOD2 &NGC~4494& S\'{e}rsic n=4 & 0 &$1.00\pm^{0.00}_{0.01}$&$0.97\pm0.01$&$0.80\pm0.01$\\
MOD2 &NGC~4494& S\'{e}rsic n=4 &0.065 &$1.00\pm^{0.00}_{0.01}$&$0.97\pm0.01$&$0.80\pm0.01$\\

MOD1 &M~32& Hernquist & 0 &$1.00\pm^{0.00}_{0.01}$ &$1.00\pm^{0.00}_{0.01}$&$0.98\pm0.01$\\
MOD1 &M~32& Hernquist & 0.0025 &$1.00\pm^{0.00}_{0.01}$&$1.00\pm^{0.00}_{0.01}$&$0.98\pm0.01$\\
MOD1 &M~32& S\'{e}rsic n=4 & 0 &$1.00\pm^{0.00}_{0.01}$&$1.00\pm^{0.00}_{0.01}$&$0.99\pm0.01$\\
MOD1 &M~32& S\'{e}rsic n=4 &0.0025 &$1.00\pm^{0.00}_{0.01}$&$1.00\pm^{0.00}_{0.01}$&$0.99\pm0.01$\\
\hline
\hline
\multicolumn{2}{|c|}{Tidal Disruption and Dynamical Friction only}\\ 
\hline
MOD4 & NGC~4889 & Hernquist & 0 &$0.37\pm0.01$ &$0.10\pm0.01$ &$0.06\pm0.01$\\
MOD4 & NGC~4889 & Hernquist & 20 &$0.42\pm0.01$ & $0.13\pm0.01$ & $0.08\pm0.01$\\
MOD4 & NGC~4889 & S\'{e}rsic n=4 & 0 &$0.48\pm0.01$ &$0.17\pm0.01$ &$0.12\pm0.01$\\
MOD4 & NGC~4889 & S\'{e}rsic n=4 & 20 &$0.52\pm0.01$&$0.20\pm0.01$&$0.14\pm0.01$\\
\hline
\hline
\end{tabular} 
\captionsetup{format=plain,labelsep=period,font={small}}
\caption{In this table the fraction of destroyed clusters (after 10 Gyr of evolution) defined by the parameter
$\chi_{\frac{R_{A}}{R_{H}}}=\frac{N_{\scriptsize{\mbox{dest}}}(t_{\tiny{\mbox{end}}})}{N_{\scriptsize{\mbox{GC}}}(t_{0})}|_{\frac{R_{A}}{R_{H}}}$
are listed. Their Poisson errors were rounded upwards. In the lower part of the table results from computations without energy-equipartition driven evaporation and without SEV are presented. 
In this way the SMBH contribution to the \textit{TDDP} in very massive and extended galaxies like NGC~4889 is demonstrated.}  
\label{GCDIS}
 \end{center}
\end{table*}
\subsubsection{Tidal Disruption Dominated Phase}\label{subsubsec:TDDP}
We want to emphasize the strong chronological aspect in the evolution of whole globular
cluster systems which can be observed in our computations (Fig.~\ref{Erosionplot}). Significant numbers of GCs are being torn apart early on,
i.e within a few crossing timescales of the galaxy at its half mass radius, $T_{\scriptsize{\mbox{cross}}}=42.26\left(R_{H}^{3}/M_{\scriptsize{\mbox{GAL}}}\right)^{0.5}
\mbox{Myr}<<T_{\scriptsize{\mbox{Hubble}}}$, which is $T_{\scriptsize{\mbox{cross}}}=\unit{3.3}{\mbox{Myr}}$ for M~32 and $T_{\scriptsize{\mbox{cross}}}=\unit{187}{\mbox{Myr}}$ for NGC~4889.
This can be seen in form of the steeply rising slope of the fraction of destroyed clusters at very early times. Hence, we characterize it as the \textit{tidal disruption dominated phase (TDDP)}.
In isotropic galaxy models (MOD2-MOD4) with central SMBHs and S\'{e}rsic $n=4$ density profiles, approx. $10\%$ (MOD4) to $40\%$ (MOD2) of all GCs are destroyed
within the \textit{TDDP}. During the \textit{TDDP}, tidal shocks dominate cluster dissolution processes. It is subsequently followed by a long term relaxation
driven dissolution phase in which surviving clusters lose mass more gently. The \textit{TDDP} can be explained as follows. By assuming the initial GC phase
space distribution to equal that of the stellar component of the host galaxy, significant numbers of clusters pass close to the galactic center within their first orbit.
Here tidal shocks cause rapid mass loss and destruction of GCs. The creation of central cores in the radial globular cluster distribution proceeds
rapidly (\S~\ref{sec:results_core}). Evidently, the fraction of destroyed GCs depends on the mass and size of the galaxy i.e. the tidal field.
The \textit{TDDP} is most pronounced in very compact and not so massive galaxies and less efficient in very extended galaxies. However, we note
that M~32 (MOD1) is an extremely compact galaxy and should not be regarded as representative for common dSph galaxies. We note that there are dwarf
elliptical galaxies with much larger spatial scales than that of M~32. See also Figure 2 in \cite{2008MNRAS.386..864D}. In such dEs the tidal field is much weaker 
and the fraction of destroyed GCs is strongly reduced. For comparison, we computed a dSph galaxy model with a S\'{e}rsic n=1 density profile
and isotropic velocity distribution (Figure~\ref{MgalvsNdis}). In this model we found no indication for a pronounced \textit{TDDP} but a strong
contribution from dynamical friction. It drives large amounts of GCs to the center where they would merge together and form a nuclear star cluster.

To get a closer insight into the dynamics of the \textit{TDDP}, we performed eight additional computations with more stringent
criteria ($x=0.8$ or $x=\infty$) for GC desintegration processes by tidal shocking. See the solid and dotted purple lines in
Figure~\ref{Erosionplot}). Evidently, the number of destroyed clusters during the \textit{TDDP} decreases and the overall slope rises less steeply.
However, even by using $x=\infty$ there is evidence for the occurence of a \textit{TDDP} in the most compact galaxy models MOD1. Interestingly, the \textit{TDDP}
does not change the total fraction of destroyed GCs after 10 Gyr but affects the temporal evolution/slope of desintegration processes.
In \S~\ref{sec:critical_discussion} we also critically review our assumptions of initial cluster sizes and galaxy models
which affect the strength of the \textit{TDDP}. In Section~\ref{subsubsec:SMBHs} we discuss the influence of secondary aspects like the central SMBH
and galactic density profile on the \textit{TDDP}. 

The influence of the host galaxy on the cluster disruption/dissolution rate becomes also evident if we calculate the normalized arithmetic mean radius, $\overline{R_{D}}/R_{H}$.
$\overline{R_{D}}/R_{H}$ is defined to be the averaged radius at which GCs in our computations were assumed to be destroyed, either by tidal shocks or relaxation driven dissolution.  
$\overline{R_{D}}/R_{H}$ anti correlates with the mass and size of the host galaxy and is largest in the compact M~32-like galaxy ($\overline{R_{D}}/R_{H}\approx1.1$) and lowest 
in the most massive and extended galaxy, NGC~4889 ($\overline{R_{D}/R_{H}}\approx0.15$). 

The existence of a rapid phase in the evolution of GCs might be of strong relevance for the fast build-up of a galaxy's field-star population from eroding clusters.
Furthermore, the existence of the \textit{TDDP} might be relevant for SMBH growth processes in the very early universe as some fraction of the debris might enter loss cone
trajectories and contribute to the feeding of the central black holes. Especially, as the majority of cluster debris is gravitationally unbound with respect
to the black hole potential and gravitational focussing would enlarge its geometric cross section.
Interestingly, the phase space distribution of the field stars originating from such a \textit{TDDP} 
should be complementary to the phase space distribution of the surviving globular clusters which is discussed in \S~\ref{sec:results_core}.

\subsubsection{Radial Anisotropy}\label{subsubsec:radialanisotropy}
\begin{figure}
\centering
\includegraphics[width=8.5cm]{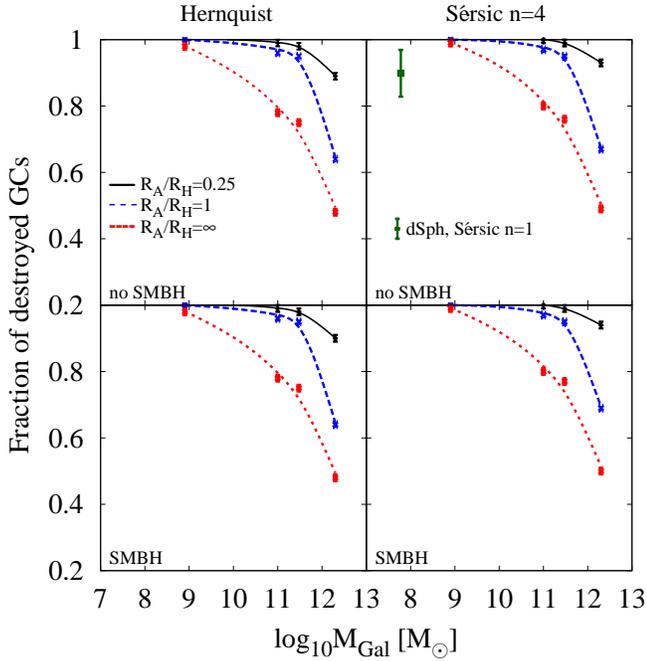}
\captionsetup{format=plain,labelsep=period,font={small}}
\caption{The fraction of destroyed clusters versus galaxy mass. The individual data points are listed in Table~\ref{GCDIS}. The colored lines
are obtained by a weighted cubic spline approximation. The fraction of eroded GCs decreases with increasing galaxy mass. For comparison,
we also computed a dSph galaxy model ($M_{\scriptsize{\mbox{GAL}}}=\unit[6\cdot10^7]{\msun}$, $R_{H}=\unit[500]{pc}$) with a S\'{e}rsic n=1 density
profile and an isotropic velocity distribution (green data point, right upper panel). The erosion rate after 10 Gyr in a dSph galaxy is strongly
reduced compared to the more compact M~32 like models. We ran computations with and without dynamical friction. We then used the mean value for
the fraction of destroyed clusters since DF has a substantial influence on the erosion rate in this model and the neglection of
DF backreaction effects on the stellar density profile (\S~\ref{sec:critical_discussion}, point vi) could lead to erroneous results.} 
\label{MgalvsNdis}
\end{figure}
The overall fraction of destroyed globular clusters in spherical galaxies with an isotropic velocity distribution (and no
central SMBH) depends on the mass and scale of a galaxy (Table~\ref{GCDIS}). While up to 100\% of all GCs are destroyed in
compact dwarf galaxies like M~32, and $75-80\%$ in mid-size galaxies, no more than $50\%$ are eroded over the course of
10 Gyr in the most massive and extended galaxies like NGC~4889 (MOD4). In Figure~\ref{MgalvsNdis} the total fraction of dissolved GCs
is plotted as a function of the mass of the galaxy. As can be seen, the initial orbital anisotropy has a considerable impact on the overall globular cluster erosion
rate in massive elliptical galaxies. Compared to the specific isotropic galaxy model MOD4 with a S\'{e}rsic $n=4$ and central SMBH, orbital anisotropy increases the fraction of 
destroyed clusters from $50\%$ ($R_{A}/R_{H}=\infty$) to $70\%$ ($R_{A}/R_{H}=1$) and $95\%$ ($R_{A}/R_{H}=0.25$).
Different formation or merger histories, and thus different degrees of radial anisotropy, may therefore be a reason for considerable scatter in the total number of 
surviving GCs in observed elliptical galaxies of similar size and mass.

The fraction of eroded GCs in compact dwarf elliptical galaxies like M~32 (MOD1) centers around 100\%. This number is insensitive to the 
initial velocity distribution. Our computations naturally explain the absence of globular clusters around M~32. However, early GC
stripping by M~31 might have occurred as well.

\subsubsection{Density Profile and SMBHs}\label{subsubsec:SMBHs}
Secondary aspects like the density profile or central SMBH exert their action only in very massive galaxies.
On average the absolute erosion rate is 1-4\% higher in the centrally more peaked S\'{e}rsic $n=4$ models.
The strongest impact is observed in the galaxy models MOD4 (Table~\ref{GCDIS}). These differences can be explained by a higher 
initial number density of GCs inside the centrally more concentrated S\'{e}rsic $n=4$ models and a steeper
gradient of the tidal field. 

The impact of SMBHs on the overall GC erosion rate after 10 billion years of evolution is insignificant.
The increase of the total destruction rate compared to models without central SMBHs does not exceed the one percent level. This is
of the same order as the assumed Poisson error related to statistics. However, this does not mean that SMBHs 
do not contribute to the exact sequence of GC dissolution processes inside galactic nuclei. It is irrelevant for the overall GC erosion rate (after 10 Gyr) 
if clusters were eroded continuously or disrupted by the central SMBH during a singe close passage. In order to isolate the impact of SMBHs during
the \textit{tidal disruption dominated phase}, computations without relaxation driven mass-loss and SEV were performed (lower part of Table~\ref{GCDIS}).
Evidently, the $\mbh=\unit{2\cdot10^{10}}{\msun}$ ultramassive black hole inside the reference galaxy MOD4,
which has similar physical properties like the BCG NGC~4889, contributed significantly to the number of tidal disruptions. 

Secondary aspects like the density profile or SMBH might also become relevant in low density dwarf or irregular galaxies in which the overall
gradient of the potential as well as the fraction of tidally disrupted GCs are small.

\subsubsection{Dynamical Friction}\label{subsubsec:DF}
\begin{figure}
\centering
\includegraphics[width=8.5cm]{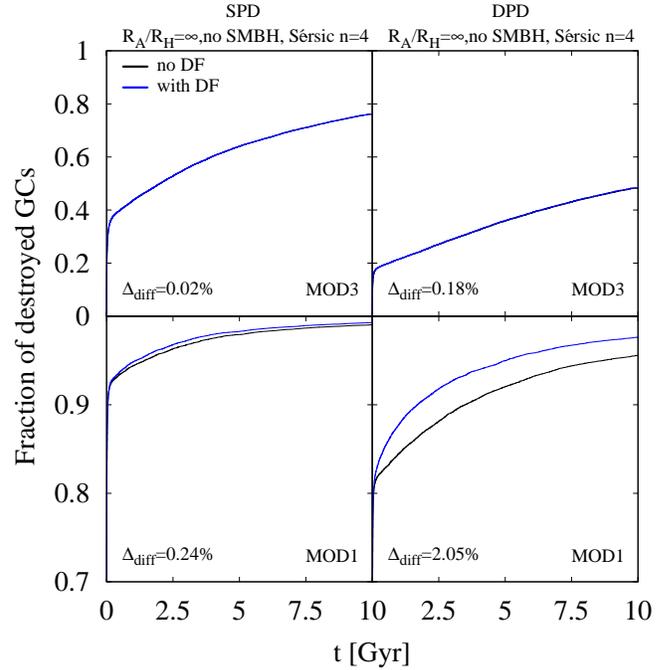}
\captionsetup{format=plain,labelsep=period,font={small}}
\caption{The fraction of destroyed clusters depends on the galaxy model, the initial cluster mass distribution (i.e. single (SPD) vs double power-law distribution (DPD)) and dynamical
friction. The parameter $\Delta_{\scriptsize{\mbox{diff}}}$ quantifies the difference in the fraction of destroyed GCs after 10 Gyr between models including DF and those where it was ignored.} 
\label{GCDIS_DPD}
\end{figure}
The influence of dynamical friction on the overall destruction rate in very massive elliptical galaxies is almost negligible. 
The reasons behind the sub-dominant impact of DF on the GC destruction rate are three-fold:
\begin{enumerate}
 \item GC masses are distributed according to a single power law GC mass function (\S~\ref{subsec:init_cond_gc_mass_size}). Most initial clusters masses
       are located at the low mass end of this distribution. Initial SEV further decreases their masses. However, the strength of de-acceleration
       by DF is proportional to cluster masses (Eq.~\ref{f3} \& \ref{f4}). Figure~\ref{GCDIS_DPD} compares the influence of DF in models with a single and a double
       power law initial cluster mass distribution but otherwise identical physical parameters. In the latter case, DF has a stronger influence on the overall GC
       erosion rate due to GCs being preferentially more massive. We chose the threshold mass, $m_{\scriptsize{\mbox{TH}}}=\unit{2\cdot10^{5}}{\msun}$, with
       slopes $\beta=0.2$ below and $\beta=2$ above $m_{\scriptsize{\mbox{TH}}}$.
 \item Low mass clusters are particularly susceptible for relaxation driven mass loss. In this way, their masses are continuously decreased so that dynamical friction gets less important
 \item Finally, the strength of DF is proportional to the distance dependent Coulomb logarithm (Eq.~\ref{f4b}) which becomes zero at small galactocentric
       distances.
\end{enumerate}       
The influence of DF on the overall destruction rate in models with a single power law cluster mass function is only evident (up to the percentage level)
in computations representing the dwarf compact elliptical galaxy M~32 (MOD1). But even in this galaxy, the tidal field strongly dominates GC erosion processes. 
The only exception where DF contributed significantly to GC erosion processes was observed in the dSph galaxy model (Figure~\ref{MgalvsNdis}).

\subsection{GC Core Formation in Giant Elliptical Galaxies}\label{sec:results_core}
\begin{figure*}
\centering
\includegraphics[width=14cm]{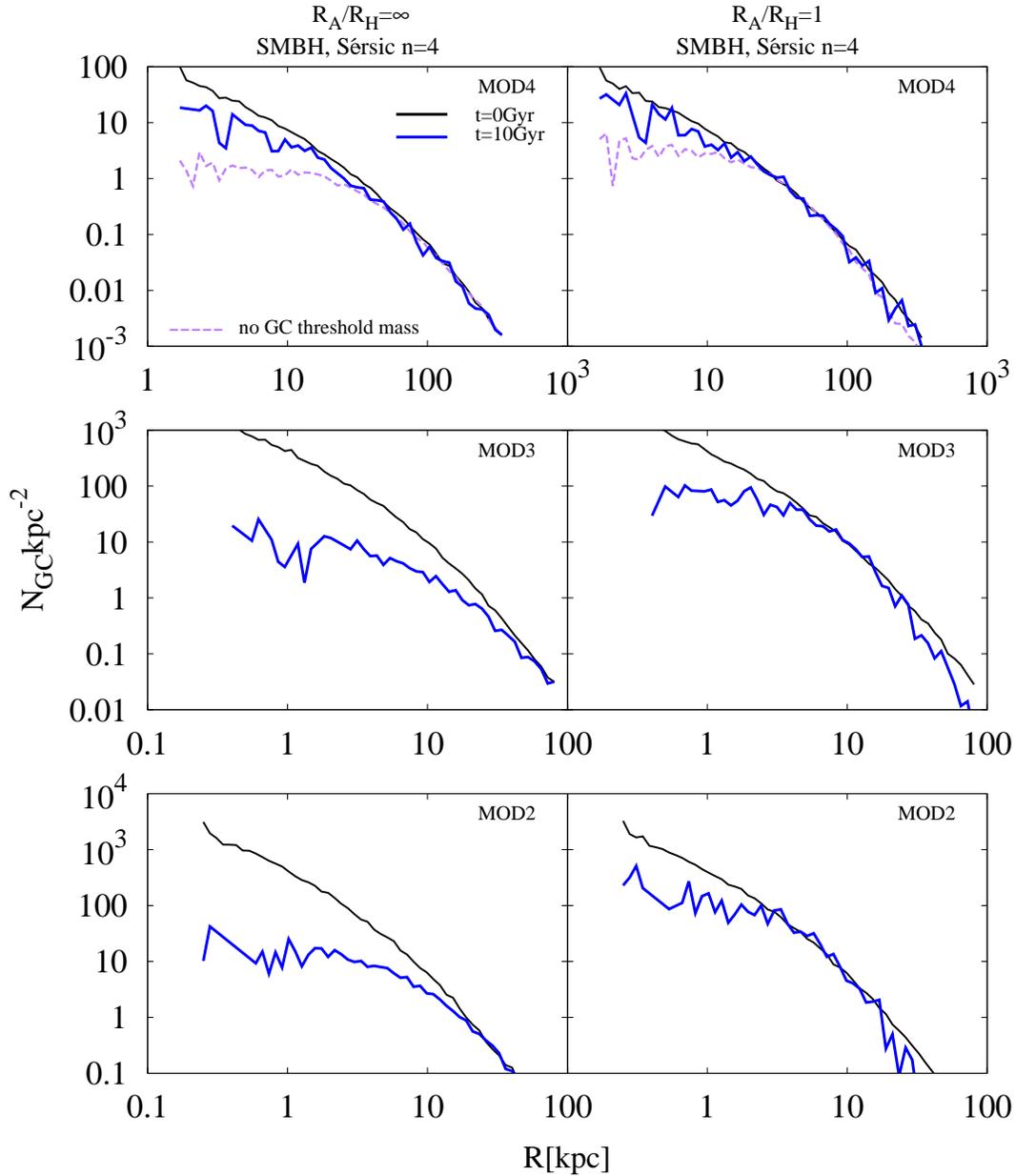}
\captionsetup{format=plain,labelsep=period,font={small}}
\caption{Projected 2D number density profiles of globular cluster systems with isotropic and radially biased ($R_{A}/R_{H}=1$) velocity distributions.
The slope of the initial GC configuration (solid black curve) also corresponds to the stellar light profile of its host galaxy. In all cases surviving globular
cluster systems turn into centrally shallower configurations. Interestingly, central core profiles seem to be less pronounced in radially biased configurations
provided they are compared to initial profiles (black lines). In anisotropic velocity configurations, GCs at large spatial scales are eroded more efficiently  
and the shape of the number density profile is less affected. Observational limitations are handled by using a GCs threshold mass
($M_{\scriptsize{\mbox{th}}}=\unit{2\cdot10^{5}}{\msun}$ in MOD4, $M_{\scriptsize{\mbox{th}}}=\unit{10^{4}}{\msun}$ in MOD3 and MOD2). Core sizes become more pronounced
when all clusters are considered (purple curves with $M_{\scriptsize{\mbox{th}}}=\unit{10^{2}}{\msun}$). However, core sizes in MOD3 and MOD2 
are close to the unfiltered values as the number of GCs below $\unit{10^{4}}{\msun}$ is negligible (see the cluster mass distribution in Fig.~\ref{GCmeanmass}). We note that for comparison 
issues the blue lines are rescaled by constant factors.} 
\label{GCProfiles}
\end{figure*}
Galaxy observations reveal the spatial globular cluster distribution to be centrally less peaked than that of the stellar light profile
\citep{1979ARA&A..17..241H, 1996ApJ...467..126F,1999AJ....117.2398M, 2009A&A...507..183C}. In this section the formation of core
profiles as a consequence of globular cluster disintegration processes in tidal fields will be investigated. The initial and final (after 10 Gyr)
2D number density profiles are compared relative to each other. This is done for the representative galaxies MOD2 (NGC~4494),
MOD3 (IC~1459) and MOD4 (NGC~4889) in our sample (\S~\ref{subsubsec:init_cond_gc_phasespace}). The initial population of 20,000
globular clusters was distributed according to the phase space distribution of S\'{e}rsic $n=4$ models with an isotropic and a radially biased
$R_{A}/R_{H}=1$ velocity distribution and a central SMBH. The results are shown in Fig.~\ref{GCProfiles} from which the
following conclusions can be drawn:\\
\begin{figure}
\centering
\includegraphics[width=8.5cm]{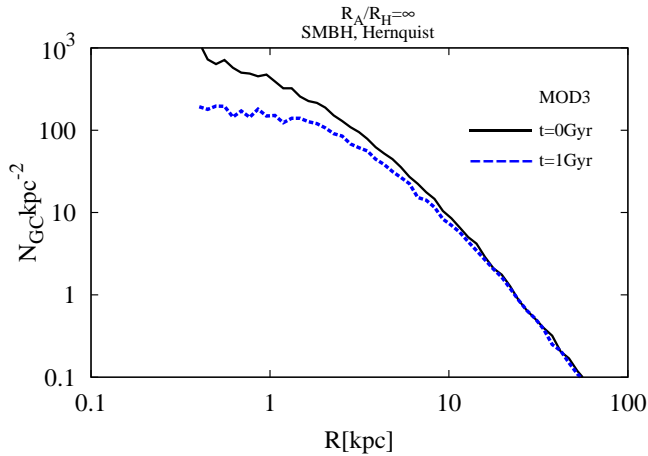}
\captionsetup{format=plain,labelsep=period,font={small}}
\caption{The central flattening of the GC number density profile progresses rapidly during the \textit{TDDP}. The blue dashed line
corresponds to the GC number density profile after one billion years of evolution. In order to emphasize the contribution from tidal shock driven cluster dissolution, mass loss through relaxation 
was neglected in this computation. The initial profile follows a Hernquist model.} 
\label{IC1459ra1e30_td2}
\end{figure}
\begin{figure*}
\centering
\includegraphics[width=14cm]{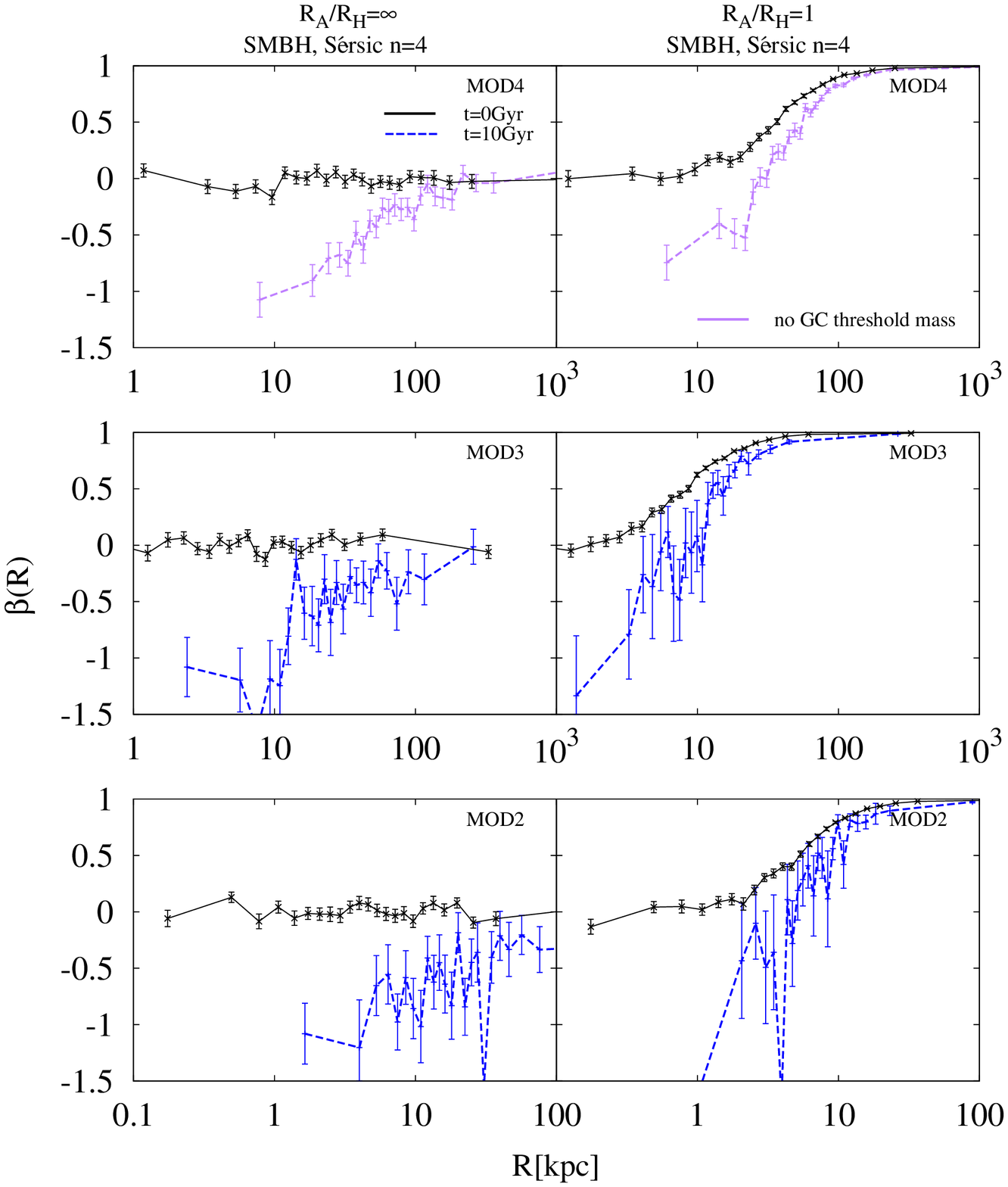}
\captionsetup{format=plain,labelsep=period,font={small}}
\caption{Final (3D) anisotropy profiles of the surviving GC population, $\beta=1-(\sigma_{\theta}^{2}+\sigma_{\phi}^{2})/2\sigma_{r}^{2}$, as a function of the galactocentric radius.
The error values are obtained by the bootstrapping method. All GCs are weighted
equally without discrimination of their mass/luminosity. In all computed models the central velocity distribution becomes tangentially biased.
In configurations with initial radial anisotropy, the tangential biased region develops within the 3D half mass radius. For MOD2 and MOD3 only clusters above the mass threshold 
$m_{\scriptsize{\mbox{GC}}}=\unit{10^{4}}{\msun}$ are taken into account (blue dashed curves). Unfiltered values are plotted in MOD4.} 
\label{betaProfiles}
\end{figure*}
\begin{enumerate}
\item In all galaxies the central globular cluster distribution becomes flattened by erosion processes.
      The outer GC number density profiles in isotropic distributions remain intact and it follows that GC distributions around the most massive and
      largest galaxies should have preserved their initial conditions. These results are in agreement with findings by \cite{2000MNRAS.318..841V}.           
      Cores are more extended in isotropic velocity distributions despite reduced numbers of destroyed clusters.
      The reason behind this apparent contradiction is related to the existence of the larger number of GCs on eccentric orbits in radially biased velocity configurations.
      GCs with large galactocentric distances also get close to the galactic center, where (at least) the less massive globular clusters are efficiently eroded.
      In this way clusters all along the radial distribution become affected over time while the overall shape (i.e. slope) of the number
      density distribution is conserved. This observation is also consistent with earlier studies (e.g. \citealt{2003ApJ...593..760V}, their figure~5).
      However, in velocity distributions of Osipkov-Merritt type with the most extreme value $R_{A}/R_{H}=0.25$, mean
      pericentric distances of GC orbits decrease with increasing galactocentric distance, resulting in a steepening of
      the outer slope of the GC distribution. This effect is discussed in more detail in Appendix~\ref{ERA}.
      
      These observations have a profound impact for the study of GC systems. Efforts to compute the number of eroded GCs by simply integrating the
      central number deficit of clusters by comparison to the stellar light component might be biased. The inferred values should be corrected for the
      influence of radial anisotropy, mass and scale of the host galaxy.
      
      A systematic study in which core sizes are obtained for comparison issues with actual data of real galaxies must also include a threshold mass
      scale in order to mimic observational limitations. We found that core sizes depend on the imposed threshold mass, $M_{\scriptsize{\mbox{th}}}$.
      This is related to the fact that massive GCs are less affected by tidal fields. The effect is shown in Fig.~\ref{GCProfiles} for the case of MOD4. Here the fraction of surviving GCs is largest and the effect is most pronounced.
      The purple dashed lines represent the unfiltered number density profiles ($M_{\scriptsize{\mbox{th}}}=\unit[10^{2}]{\msun}$) whereas the blue lines correspond to GCs in excess of $M_{\scriptsize{\mbox{th}}}=\unit{2\cdot10^{5}}{\msun}$.
      This corresponds to the detection limit at the distance of NGC~4889. However, slope differences between unfiltered and those with $M_{\scriptsize{\mbox{th}}}=\unit{10^{4}}{\msun}$ are small
      in MOD2 and MOD3, owing to a negligible fraction of GCs below $\unit{10^{4}}{\msun}$ (Fig.~\ref{GCmeanmass}). 
      The initial profiles (black solid lines) were rescaled to match the final profiles at large galactocentric radii.
      We take from this figure that the observation of a mass dependent core size of a globular cluster system might be proof of cluster dissolution as the origin of the core (\S~\ref{sec:intro}).
\item Core formation occurs on short cosmological timescales. Shortly after the \textit{TDDP} the GC number density 
      profiles are centrally flattened (Fig.~\ref{IC1459ra1e30_td2}).
\item Despite the influence of radial anisotropy, cores are pronounced in less massive galaxies, as here the percentage of disrupted clusters as well as the ratio
      $\overline{R_{D}}/R_{H}$ is highest. However, spatially extended cores are also found in the most luminous galaxies in the universe,
      and our computational results indicate that observed GC profiles with central cores \citep{1996ApJ...467..126F} are created by disruption and dissolution processes.
      To see if our computations are in agreement with observations, we plotted GC number density profiles for NGC~4889 (MOD4, blue lines) by taking observational limitations into account.
      For a typical mass-to-V-band light ratio $\mathnormal{\Upsilon}_{V}=1.5$ \citep{2005ApJS..161..304M}, $m_{\scriptsize{\mbox{GC}}}=\unit{2\cdot10^{5}}{\msun}$ corresponds to the threshold mass below a
      GC would be undetected at the distance of $D\approx\unit{100}{\mbox{Mpc}}$ \citep{2009AJ....137.3314H}.
      As can be seen in Fig.~\ref{GCProfiles}, central flattening is compatible with our models within the central few kiloparsecs. This is in agreement 
      with the inner parts of the galaxy $r\approx\unit{3.5}{\mbox{kpc}}$ (figure 6 in \citealt{2009AJ....137.3314H}) where the GC number density profile becomes more shallow than the stellar light profile of NGC~4889. 
      However, in a more realistic scenario, erosion is only partly responsible for the observed central shallow profiles, as in these galaxies' major merger events will
      also contribute to the spatial flattening of the radial GC density profiles \citep{2006A&A...445..485B}. 
\item Our computations also reflect the preferential destruction of GCs on elongated orbits and the consequences
      for the dynamics of the surviving globular cluster system. After 10 Gyr of evolution, the central regions of the plotted models (Figure~\ref{betaProfiles}) show strong
      signs of a tangential bias subject to the preferential survival of GCs on circular orbits. In models with initial radial anisotropy, a tangentially biased region develops within the 
      3D half mass radius. At large distances the radial anisotropy is reduced but still persists at significant levels. 
\end{enumerate}
Our computations demonstrate a relation between core sizes of globular cluster systems and the host galaxies mass and velocity distributions.
A quantitative evaluation of this correlation will be an interesting task for a follow-up investigation. 

\subsection{Final Globular Cluster Mass Distribution}\label{sec:mean_mass}
\begin{figure*}
\centering
\includegraphics[width=14cm]{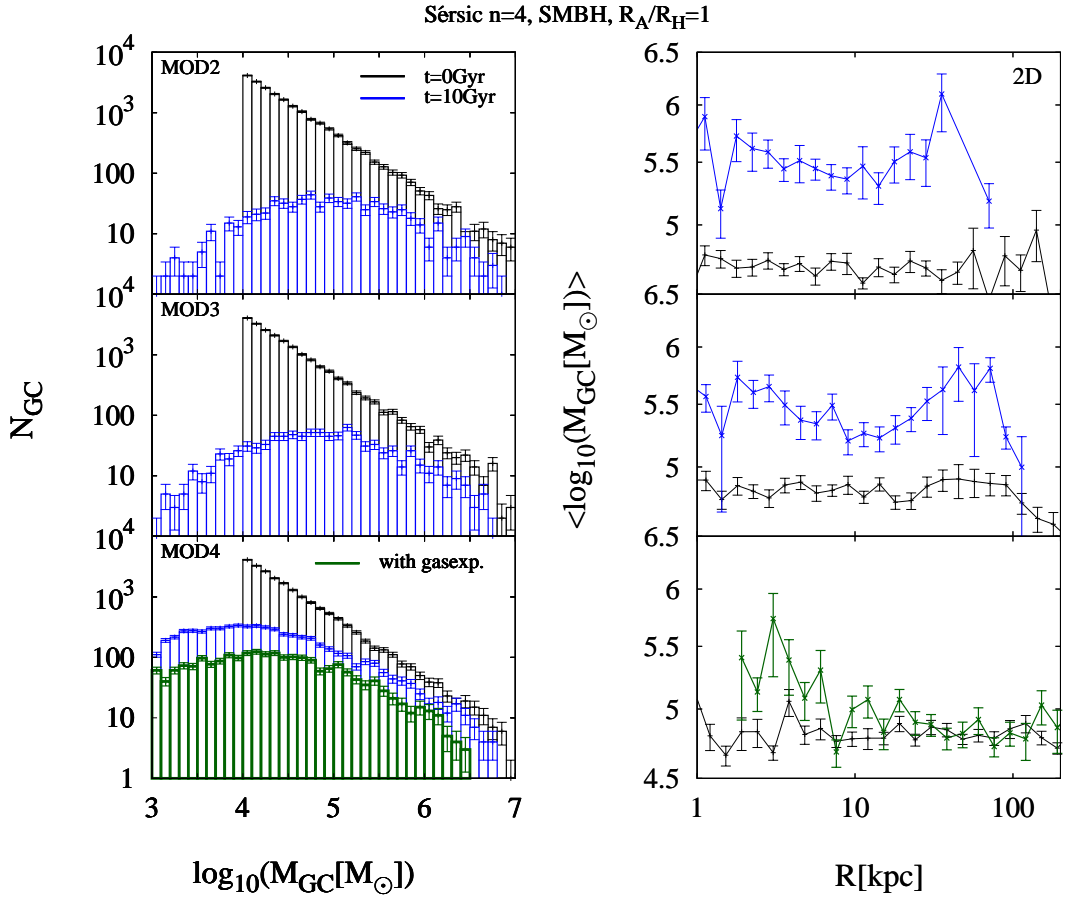}
\captionsetup{format=plain,labelsep=period,font={small}}
\caption{Left panels: Initial (black) and final (blue histograms) globular cluster mass functions. Error bars show Poisson uncertainties.
A moderate degree of radial anisotropy ($R_{A}/R_{H}=1$) transforms initial power law distributions into bell shaped curves (MOD2 and MOD3).
The inclusion of a primordial gas expulsion phase (green histograms and curve, MOD4) contributes to the shaping of a bell shaped mass function even in the most massive galaxies. 
Right panels: The globular cluster mean mass versus galactocentric distance obtained for 2D radial binning.
Uncertainties were derived via bootstrapping. Despite a large degree of scatter MOD2 \& MOD3 are compatible with a constant mean mass over a broad range of galactocentric distances (see text for more details).
The mean mass centers around $(3-4)\cdot\unit{10^{5}}{\msun}$ and is strongly affected by a small number of GCs more massive than a few $\unit{10^{6}}{\msun}$.
By assuming a Schechter-type function for initial cluster masses with a particular turn-down mass in the high GC mass regime, the final mean mass might naturally
lower to $\unit{2\cdot10^{5}}{\msun}$. At least in spiral galaxies in which the initial cluster mass distribution can be observed directly, it is compatible with such a
Schechter mass function \citep{2009A&A...494..539L}.} 
\label{GCmeanmass}
\end{figure*}
Fig.~\ref{GCmeanmass} shows the evolution of the globular cluster mass function (left-hand panels).
Results are plotted for three galaxies with S\'{e}rsic $n=4$ density profiles, central SMBHs and a radially biased velocity distributions.
Evidently, a moderate degree of radial anisotropy ($R_{A}/R_{H}=1$) transforms initial power law distributions into bell shaped curves (upper and middle panels)
with a peak at approx. $\unit{10^{5}}{\msun}$. However, the GC destruction rate in the most massive and extended galaxies (MOD4) is reduced owing to weak tidal fields.
Here, the total fraction of dissolved GCs is not high enough to turn a power law distribution into a bell shaped curve peaking at $\unit{2\cdot10^{5}}{\msun}$. Stronger initial anisotropy or mass loss
related to gas-expulsion \citep{2002MNRAS.336.1188K, 2008MNRAS.384.1231B} might represent one solution to this discrepancy. Indeed, including
gas expulsion during the gas rich cluster phase results in a bell shaped mass function after 10 Gyr of evolution. This is shown in Fig~\ref{GCmeanmass} in
form of green shaded histograms. In order to mimic the effects of gas expulsion on the embedded cluster mass function, we applied Eq. 6,8,9 from \cite{2002MNRAS.336.1188K}.
However, even by considering gas expulsion, the distribution peaks at a few $\unit{10^{4}}{\msun}$ instead of $\unit{2\cdot10^{5}}{\msun}$. An additional effect which might
naturally explain this discrepancy in the most massive and extended galaxies is related to the idea, that BCGs are partly grown from galaxy mergers of smaller 
constituents at high redshift. Furthermore, the progenitor galaxies of the most massive ones today were initially much more compact
(see e.g.~\citealt{2007MNRAS.382..109T}, \citealt{2008ApJ...677L...5V}). Within these progenitor galaxies, the cluster mass functions quickly transformed into a bell
shaped form before they merged together to form the BCG. Notice the fast temporal evolution in Fig.~\ref{Erosionplot} in the less massive but more compact galaxies.

The relations between GC mean masses and galactocentric distances are plotted in the right-hand panels of Fig.~\ref{GCmeanmass}. Despite a large degree
of scatter due to low number statistics of surviving clusters, models MOD2 \& MOD3 are in agreement with the hypothesis of having a constant GC
mean mass over a broad range of galactocentric distances. The residual slope $b$ which is obtained from a linear regression in the
intervall $r \in [1-200]$ kpc is of the order of the one sigma error bar ($b/\Delta b=0.98$, MOD2) and ($b/\Delta b=1.1$, MOD3)
respectively. However, the hypothesis of a constant GC mean mass over a large galactocentric distance is rejected for MOD4 ($b/\Delta b=2.8$).
Our results imply that in order to reproduce observed GC properties of BCGs, their merger history from more compact progenitors should be considered.

Although our computations were not designed to reproduce GC characteristics of particular galaxies but instead to illustrate systematics,
they already reproduce a lot of observed features: a bell shaped mass distribution, a nearly constant globular cluster mean mass over large galactocentric distances and
shallow central number density profiles.   

\section{CRITICAL DISCUSSION AND OUTLOOK FOR FUTURE WORK}\label{sec:critical_discussion}
In this Section we will critically review our assumptions and results. This will ease the efforts to identify potential weaknesses
and help to improve follow-up studies.
 \begin{enumerate}
  \item The specific implementation of tidal shock and their relevance for disruption processes (\S~\ref{subsubsec:tidal_shocks})
        requires the orbital angular momentum to be conserved or decreasing due to dynamical friction. This is because a globular cluster does
        not necessarily become completely unbound by tidal shocks once the ratio of half mass radius to Jacobi radius exceeds $r_{H}/r_{J}=0.5$.
        If the angular momentum is conserved or monotonically decreasing (which is the case in spherical galaxies), such a cluster will pass the same
        or an even stronger tidal field within the next crossing timescale until it would become eroded. In all computed galaxies the vast majority of crossing timescales
        is significantly below the total duration of the integrations, thus yielding safe lower limits on the number of disruptions.
        A more complicated situation emerges in galaxies deviating from spherical symmetry due to the existence of trajectories where the
        directional components of the angular momentum vector change in time. In such galaxies, a GC passing a region in which $r_{H}/r_{J}=0.5$
        might not repeat doing so for a long time. The criterion for disruption processes by tidal shocks in non-spherical galaxies represents a much
        more challenging task and will be part of future studies. We also note that our disruption criterion was adjusted by means of
        direct \textsc{Nbody6} integrations in one particular galaxy model as well as by using one particular cluster model. However, in order to compensate 
        these shortcomings we changed the parameter $x$ to even higher values than $x=0.5$ and discussed the systematics. We found no quantitative differences in the outcomes.
        In addition to that our SEV and relaxation driven dissolution implementation (\S~\ref{subsubsec:dyn_of_GSs_relaxation}) was calibrated in
        direct $N$-body computations \citep{2003MNRAS.340..227B} which are based on a Kroupa IMF with lower and upper mass limits $\unit[0.1]{\msun}$ and $\unit[15]{\msun}$.
        The (initial) mass loss through stellar evolution would increase by using a higher upper mass limit. However, this would mostly affect the initial correction factor which has 
        no influence on the shape of the single power law GC mass distribution (\S~\ref{subsec:init_cond}).
  \item In our computations GC half mass radii were distributed according to relation Eq.~\ref{f10} and were then
        integrated by leaving their sizes unchanged. The strength of tidal shocks and hence the efficiency of tidal disruption processes depends on the compactness (i.e size)
        of globular clusters. Therefore, the percentage of disrupted GCs during the \textit{tidal disruption dominated phase} depends on initial cluster sizes. 
        If GCs would be much more compact after their rapid gas expulsion phase, the impact of the \textit{TDDP} on the overall cluster erosion rate would be reduced.
        In future studies more realistic initial conditions as well as cluster size evolution should be included. However, the same criticism also applies to
        the used galaxy models which in this study were assumed to be non-evolving. \cite{2008ApJ...677L...5V} show that massive elliptical galaxies at high
        redshifts were more compact than today. In more compact progenitor galaxies, the \textit{TDDP} on the other hand would be very pronounced
        and might quickly transform a single power law cluster mass function into a bell shaped form. After (dry) merging processes these galaxies will
        inflate their sizes but the bell shaped cluster mass function should remain unaffected. In conclusion, the efficiency of the \textit{TDDP} depends
        on cluster and galaxy size evolution. 
  \item Massive (elliptical) galaxies display a bimodal color distribution of globular clusters which have different metallicities,  
        kinematics and number density profiles \citep{1993MNRAS.264..611Z, 1997AJ....113.1652F, 2006ARA&A..44..193B, 2012MNRAS.425...66F}. These
        cluster populations are leftovers of different star-formation events. The red and metal rich GC population is centrally more concentrated and
        follows the stellar light profile of its host galaxy, whereas the number density profile of metal poor GCs is flatter and dominates the GC system at large distances. 
        Our computations address the evolution of the GC distribution which traces the galaxy light and we neglect GC populations which were formed (or accreted) later on.
  \item The compact dwarf galaxy M~32 does not contain any globular clusters. Our computations indicate that they might have been eroded in the strong tidal field of this galaxy. 
        The real stellar density profile of M~32 deviates at distances below 15 arcsec ($\approx \unit[55]{pc}$) and above 100 arcsec ($\approx \unit[370]{pc}$) from 
        a S\'{e}rsic $n=4$ profile \citep{1987AJ.....94..306K} which we used in our computations. The central density inside M~32 is even higher than the
        corresponding density of a $n=4$ profile (see Figure 4 in \citealt{1987AJ.....94..306K}). This would result in an even stronger tidal field and 
        an increase of the actual disruption rate. Therefore our results concerning the erosion rate in compact M~32 like dwarf galaxies with Hernquist or S\'{e}rsic $n=4$ density profiles
        should hold for M~32 itself. However, these results should not be applied to ``more common'' dwarf spheroidal galaxies (dSph) which are less massive, less dense and
        which have shallower density profiles (e.g. $n=1$). The GC erosion rate in dSphs is reduced as indicated by the one computation with a S\'{e}rsic $n=1$ density profile (Figure~\ref{MgalvsNdis}).      
  \item As already mentioned our computations are governed by the stellar density profiles specified in \S~\ref{subsubsec:init_cond_gc_phasespace}. The next logical extension
        would be to use the cumulative density profiles from the stellar, dark matter and gas component. It has to be investigated whether the extended
        isothermal density profiles of DM halos would significantly alter the GC erosion rate which is dominated by tidal effects deep within the galaxy where the stellar density dominates. 
  \item Dynamical friction (\S~\ref{subsubsec:df}) is implemented as an external routine in the \textsc{Muesli} code. While this is a commonly used 
        strategy in numerical investigations, care has to be taken. By assuming GCs to be immune to dissolution processes, all of them would accumulate 
        within given time periods near the center of the galaxy, driving the mass density upwards. In reality, DF is an energy conserving process and 
        while compact objects spiral inwards, stellar mass is driven outwards. These back-reaction effects are not considered in this study. However, due to 
        the sub-dominant role of DF in our computations, back-reaction processes will have a minor impact on the inferred results.
  \item The main focus of this paper is about destruction rates of GCs by tidal shocks and relaxation driven mass loss in tidal fields of
        spherically symmetric galaxies. We kept it simple and neglected the fate of dissolving GCs and how their debris might affect internal dynamics of galaxies,
        e.g. by forming a nuclear star cluster \citep{1975ApJ...196..407T, 2011ApJ...729...35A, 2013ApJ...763...62A, 2013arXiv1308.0021G}. These issues
        as well as direct SMBH loss cone studies will be part of later studies. To handle them with our \textsc{Muesli} code requires detailed
        understanding of GC dissolution mechanisms in evolving galaxies. Nevertheless, our computations already indicate a chronological
        aspect in the erosion of globular cluster systems which might be of relevance for the fast build-up of massive black holes
        in the early universe.
 \end{enumerate}

\section{Conclusions}\label{sec:conclusion}

We developed a versatile code, named \textsc{Muesli}, designed to investigate the dynamics and evolution of globular cluster systems in elliptical galaxies.
It uses the self-consistent field method (SCF) with a time-transformed leapfrog scheme to integrate orbits of field stars and GCs. 
In this way, velocity-dependent forces like dynamical friction and post-Newtonian effects of central massive black holes can be handled accurately.
In order to be able to treat spherical galaxies with anisotropic velocity distributions (as well as non-spherical galaxies), the code uses the
ellipsoidal generalization of Chandrasekhar's dynamical friction formula \citep{1992MNRAS.254..466P}. The advantage of \textsc{Muesli} lies in its
flexibility to evaluate the impact of complex physical processes on the erosion rates of globular clusters (GC) in evolving galaxies. 

In a first application, we have investigated if flat central cores in GC distributions around massive elliptical galaxies
result from tidal disruption events (TDEs) and cluster dissolution processes through relaxation. Furthermore, we explored the question
if the strong tidal field within the compact dwarf galaxy M~32 is responsible for lack of GCs in this galaxy.

We used a power-law distribution for the GC masses, and set the initial phase-space distribution of the GCs equal to the stellar phase-space
distribution of the host galaxy. The rapid phase of gas expulsion was ignored with the exception of one model. We assumed two cluster dissolution channels:
(i) A slightly modified version of relaxation driven mass loss in tidal fields (which also handles SEV) from \cite{2003MNRAS.340..227B}
was implemented. Once a cluster mass becomes less than $m_{\scriptsize{\mbox{GC}}}=\unit{100}{\msun}$, it
is assumed to be dissolved by relaxation. Additionally (ii), we identified a tidal disruption criterion in terms of the ratio of cluster half-mass radius,
$r_H$, to Jacobi radius, $r_J$, in that no cluster was able to survive for a significant amount of time, when the ratio $x = r_H/r_J$ passed
a threshold of $x=0.5$. The condition for globular cluster disruption in tidal fields was calibrated by means of direct $N$-body experiments. For this purpose, we used
the star cluster code \textsc{Nbody6} to compute the evolution of massive clusters on various orbits within the tidal field of a host galaxy.

We found that, after 10 Gyr of evolution, all computed GC systems show signs of central flattening with the central core size depending
in a non-trivial way on the mass, scale and anisotropy profile of the host galaxy and threshold GC mass. Galaxies with highly radially biased velocity distributions lose a
significant fraction of clusters also at large galactocentric radii. As a result the cores, in their central density profiles are less pronounced
than in galaxies with isotropic distributions. The primary factors which determine the disruption rate of GCs are the half-mass radius and mass of the galaxy
and the initial degree of radial anisotropy of the GC system. For host galaxies with an isotropic velocity distribution, the fraction of disrupted globular clusters
is nearly 100\% in very compact, M~32-like dwarf galaxies. 

The rate is lowest in the most massive and extended galaxies (50\%) like NGC~4889.
The arithmetic mean radius, $\overline{R_{D}}$, where most GC destruction occurred during the last 10 billion years, is roughly equal to the
(3D) half-light radius $R_{H}$ in compact dwarf ellipticals and drops to $0.15R_{H}$ in massive elliptical like NGC~4889. 
An isotropic initial velocity distribution is mostly preserved at large radius ($R>R_{H}$), while the GC velocity profile close
to the galactic center become less radial or even tangentially biased. Different degrees of initial radial anisotropy may 
be the reason for a considerable scatter in the total number of GCs around more massive elliptical galaxies (see Table~\ref{GCDIS}).
In compact M~32-like galaxy models with radial anisotropy no single GC survived.

The influence of dynamical friction on the overall GC erosion rate in massive elliptical galaxies is insignificant as long as the
initial cluster mass function follows a power law distribution with slope $\beta=2$. However, DF yields a small contribution in compact dwarf ellipticals
like M~32. Secondary effects like the density profile or the presence of a central massive black hole manifest their influence
only in the most massive and extended galaxies. An ultramassive black hole with a mass above ten billion solar masses inside a galaxy
like NGC~4889 has a considerable impact on tidal disruption processes. Its presence increases the total fraction of destroyed GCs during the violent phase of
tidal disruptions by 2\% to 5\% in absolute terms.

We also found that globular cluster erosion processes result in a bell shaped GC mass function and a nearly
constant relation between GC mean mass and galactocentric distances as long as the galaxies are not too extended and radially
biased. Observations of bell-shaped GC mass functions in extended galaxies may indicate that their GC populations
were formed in more compact building blocks of these galaxies, which later merged to form the present-day host.

Finally, our results show a strong chronological aspect in the evolution of globular cluster systems. That is, most tidal
disruptions occur at early times, on dynamical timescales of the host galaxy. Hence, we call this a \textit{tidal disruption dominated phase} in the evolution
of globular cluster systems. Our simulations strongly suggest that the number of GCs in most galaxies was much higher at their formation. Therefore, depending on the fraction of stars
in a galaxy which were born in globular clusters, the debris of the disrupted
clusters should constitute a significant amount of a galaxy's field population. In the extreme case that all stars in galaxies were born in globular clusters, our study would imply
that larger galaxies like NGC 4889 have to be the merger product of many smaller galaxies and/or that the progenitor galaxies were initially much more compact because otherwise 10-50\% of its
stellar mass would still have to be locked up in globular clusters (Fig.~\ref{Erosionplot}). Given the fact that only about 0.1\% of all stars seem to be locked up in globular
clusters nowadays, our study prefers building blocks of galaxies in the early universe to either have a small fraction of stars being born in very massive globular clusters, or being relatively
compact like M\,32, or having highly radially biased GC distributions.

Interestingly, we predict the field population coming from disrupted GCs to have complementary orbital
properties to the phase space distribution of the surviving clusters. Moreover, we predict the centrally cored GC distributions around SMBHs to be tangentially biased,
and thus parts of the field star population to have a pronounced radially biased component from cluster debris. The diffusion of this cluster debris in phase space (in combination with 
gravitational focussing relevant for unbounded matter) might therefore contribute to the rapid growth of SMBHs in the early universe through the refilling of the black hole loss cone.
To which degree will be subject to a future study.

\section*{ACKNOWLEDGMENTS}
The authors would like to thank an anonymous referee for useful comments that helped to increase the quality of the manuscript, and J.P. Ostriker for stimulating discussions.
The work of this paper was supported by the German Research Foundation
(DFG) through grants KR 1635/39-1 within the programme "Constraining the dynamics and growth history of super-massive black holes (SMBHs) in the lowest and highest mass
regime", and through DFG project KR 1635/28-1. AHWK would like to acknowledge support through DFG Research Fellowship KU 3109/1-1 and from NASA through Hubble Fellowship
grant HST-HF-51323.01-A awarded by the Space Telescope Science Institute, which is operated by the Association of Universities for Research in
Astronomy, Inc., for NASA, under contract NAS 5-26555. HB acknowledges support from the Australian Research Council through
Future Fellowship grant FT0991052.


\appendix 
\section{TESTING}\label{sec:testing}
\subsection{Discreteness Noise}\label{subsec:testing_discreteness}
When the potential becomes updated in time intervals $\Delta t_{\scriptsize{\mbox{up}}}$ by the SCF algorithm,
fluctuations in the overall particle distribution give rise to irregular oscillations of the virial ratio \citep{Hernquist1992}.
Additionally, improper selection of the expansion coefficients, especially the angular expansion order leads to multipole
induced precession.

We inferred the magnitude of these fluctuations, which inversely ($\propto N^{-0.5}$) depend on the particle number,
by the computation of spherical and axisymmetric S\'{e}rsic $n=4$ and Hernquist profiles. All models were evolved forward up to
100 $N$-body time units. We varied the particle number ($N=2\cdot10^{5}$ and $N=4\cdot10^{6}$), the timescales of potential evaluation
($\Delta t_{\scriptsize{\mbox{up}}}=1$ and $\Delta t_{\scriptsize{\mbox{up}}}=100$) as well as the radial and angular
expansion order (between $l=0$ and $l=20$). In this way we were able to estimate the fluctuations and their relevance for the
accuracy of orbit integrations of elliptical galaxy models. For testing purposes (only), the axisymmetric models were generated by
simply reducing the z-components by factors of two. Without adjusting particle velocities properly, axisymmetric
models generated in such a way are dynamically unstable. The only reason for using them instead of existing virialised
axisymmetric models generated from cold collapse computations was subject to more controlled conditions required
for the evaluation of discreteness noise effects. Orbits in axisymmetric models were therefore computed in fixed potentials
i.e. by using $\Delta t_{\scriptsize{\mbox{up}}}=100$.

Ideally, a test particle with zero velocity in z-direction has to orbit the galaxy without
changing its z-component. Hence, the maximal drift of the angular momentum vector $\vec{L}$, i.e. the maximal angle $\alpha_{m}$
between $\vec{L}_{t_{0}}$ and $\vec{L}_{t_{i}}$, was used as one criterion for the accuracy of orbit integration.
As an additional criterion we have used the standard deviation $\sigma_{c}=\left(\frac{1}{N_{I}-1}\sum_{i=1}^{N_{I}}(r_{i}-\overline{r})^{2}\right)^{0.5}$
from circular motion at different galactocentric distances $r=0.1,1,10$. This quantity reveals the magnitude of potential fluctuations. 

Apparently, relaxation arising from discreteness noise is sub-dominant when performing integrations with high particle numbers
($N\geq4\cdot10^{6}$) and by using low angular expansion terms i.e. $l=0$ for the computation of spherical galaxies.
In axisymmetric or triaxial galaxies torques from the overall matter configuration are orders of magnitudes larger than
local anisotropies or discreteness noise. The integration inaccuracies are listed in the Table~\ref{tablescf}.
On the basis of tabulated data, several trends can be obtained. The amplitude of discreteness noise effects are
anti-correlated with the total number of particles. Torques induced by angular multipole expansion affect mostly trajectories
with short orbital periods while the precession diminishes altogether by using lowest order (spherical) terms.
Converging solutions of the underlying density profile are obtained with high radial order terms. For the parameter space
of our main computation, radial orbit fluctuations correspond to only a few tens of parsec over timescales of several
billion years when scaled to the proportions of giant elliptical galaxies like NGC~4889
(\S~\ref{subsubsec:init_cond_gc_phasespace}) and sub-parsec scales for the smallest galaxies.

\begin{table*}
\begin{tabular}{|c|c|c|c|c|c|c|} 
  \hline
 \bf{Hernquist}& N & $\Delta t_{\scriptsize{\mbox{up}}}$ & $n$ & $l$ & $\sigma_{c}$(r=0.1,1,10)  & $\alpha_{m}$(r=0.1,1,10) \\ 
\hline
 spherical &$2\cdot10^5$&1 & 30 & 0&0.002/0.006/0.003 & 0/0/0  \\
 spherical &$2\cdot10^5$&100 & 30 & 0&0.0001/0.0001/0.0004 & 0/0/0 \\
 spherical &$2\cdot10^5$&1 & 10 & 10& 0.008/0.01/0.003& 1.8/0.1/0.001 \\
 spherical &$2\cdot10^5$&100 & 10 & 10&0.003/0.002/0.007 & 1.8/0.08/0.002  \\
 spherical &$2\cdot10^5$& 100 & 10 & 5&0.003/0.002/0.006 & 1.9/0.07/0.0006 \\

 axis-sym. & $2\cdot10^5$&100 & 12 & 1& --- &  0/0/0\\
 axis-sym. &$2\cdot10^5$&100 & 12 & 2& --- & 0.06/0.008/0.0001 \\
 axis-sym  &$2\cdot10^5$&100 & 12 & 5& --- &  0.3/0.008/0.0007 \\
 axis-sym  &$2\cdot10^5$&100 & 12 & 10& --- & 0.1/0.02/0.001 \\
 axis-sym  &$2\cdot10^5$&100 & 20 & 20& --- &  0.09/0.03/0.003\\

 spherical &$4\cdot10^6$&1 & 30 & 0&0.0009/0.002/0.0005 & 0/0/0  \\
 spherical &$4\cdot10^6$&100 & 30 & 0&0.00001/0.00002/0.00009 &0/0/0  \\
 spherical &$4\cdot10^6$&1 & 10 & 10& -/0.005/0.0003 & -/0.01/0.0001  \\
 spherical &$4\cdot10^6$&100 & 10 & 10&0.0001/0.0003/0.0003 &   1.4/0.02/0.0003  \\
 spherical &$4\cdot10^6$&100 & 10 & 5& 0.0001/0.0002/0.0003&  1.4/0.01/0.0005 \\

 axis-sym. &$4\cdot10^6$&100 & 12 & 1&--- & 0/0/0  \\
 axis-sym. &$4\cdot10^6$&100 & 12 & 2& ---& 0.01/0.001/0.00002 \\
 axis-sym  &$4\cdot10^6$&100 & 12 & 5& ---&  0.03/0.001/0.0001 \\
 axis-sym  &$4\cdot10^6$&100 & 12 & 10&--- &  0.3/0.002/0.0001 \\
 axis-sym  &$4\cdot10^6$&100 & 20 & 20& ---&  0.3/0.003/0.0003\\

\hline
\bf{S\'{e}rsic} $n=4$ & N & $\Delta t_{\scriptsize{\mbox{up}}}$ & $n$ & $l$ & $\sigma_{c}$(r=0.1,1,10)  & $\alpha_{m}$(r=0.1,1,10) \\
\hline
 spherical &$2\cdot10^5$&1 & 30 & 0& 0.001/0.0004/0.002& 0/0/0  \\
 spherical &$2\cdot10^5$&100 & 30 & 0&0.00003/0.00005/0.0002 & 0/0/0 \\
 spherical &$2\cdot10^5$&1 & 10 & 10& 0.005/0.02/0.003& 1.7/0.03/0.002  \\
 spherical &$2\cdot10^5$&100 & 10 & 10& 0.001/0.002/0.01& 0.4/0.04/0.001  \\
 spherical &$2\cdot10^5$&100 & 10 & 5&0.001/0.002/0.01 & 0.3/0.08/0.0006 \\

 axis-sym. &$2\cdot10^5$&100 & 12 & 1& ---& 0/0/0  \\
 axis-sym. &$2\cdot10^5$&100 & 12 & 2&--- & 0.03/0.02/0.0004 \\
 axis-sym  &$2\cdot10^5$&100 & 12 & 5& ---& 0.01/0.02/0.0004  \\
 axis-sym  &$2\cdot10^5$&100 & 12 & 10&--- & 0.02/0.02/0.0005  \\
 axis-sym  &$2\cdot10^5$&100 & 20 & 20& ---& 0.02/0.02/0.002 \\

 spherical &$4\cdot10^6$& 1 & 30 & 0& 0.0003/0.0009/0.0009 & 0/0/0 \\
 spherical &$4\cdot10^6$&100 & 30 & 0& 0.000004/0.00004/0.0004& 0/0/0  \\
 spherical &$4\cdot10^6$&1 & 10 & 10& 0.002/0.002/0.02& 0.6/0.009/0.0001  \\
 spherical &$4\cdot10^6$&100 & 10 & 10&0.00007/0.0005/0.02  & 0.5/0.02/0.0003  \\
 spherical &$4\cdot10^6$& 100 & 10 & 5& 0.00007/0.0005/0.02& 0.5/0.02/0.0003 \\

 axis-sym. &$4\cdot10^6$&100 & 12 & 1&--- & 0/0/0  \\
 axis-sym. &$4\cdot10^6$&100 & 12 & 2&--- & 0.004/0.001/0.0001 \\
 axis-sym  &$4\cdot10^6$& 100 & 12 & 5& ---& 0.007/0.002/0.0004  \\
 axis-sym  &$4\cdot10^6$&100 & 12 & 10& ---&   0.008/0.004/0.0005 \\
 axis-sym  &$4\cdot10^6$& 100 & 20 & 20& ---& 0.01/0.005/0.0005 \\
\hline
\end{tabular} 
\captionsetup{format=plain,labelsep=period,font={small}}
\caption{Parameters of the discreteness noise evaluation. N specifies the total number of particles and $\Delta t_{\scriptsize{\mbox{up}}}$
is the characteristic timescale at which the overall potential becomes re-evaluated by the SCF algorithm. The parameters $n,l$ correspond to
the order of the radial and angular expansion terms. $\sigma_{c}$ measures the standard deviation from circular motion at three 
different radii (r=0.1,1,10) and $\alpha_{m}$ traces the maximal angular deviation caused by multipole induced precession at the same radial distances. It is given in radians. 
There is no dipole moment induced precession (l=1) since all particles were initially inverted (and doubled) at the origin. The main computations are performed
with five times higher particle numbers than the largest test models.}
\label{tablescf}

\end{table*}

\subsection{The Conservation of the Space Phase Distribution}\label{subsec:testing_phasespace} 
\begin{figure}
\centering
\includegraphics[width=8.5cm]{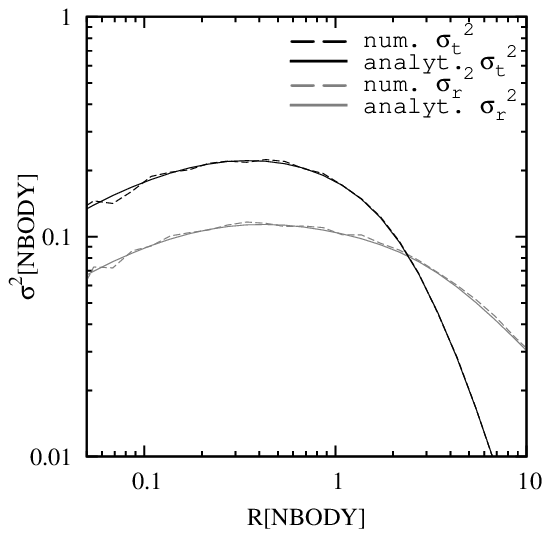}
\captionsetup{format=plain,labelsep=period,font={small}}
\caption{Radial and tangential velocity dispersion profile of an anisotropic Hernquist model ($R_{A}=R_{H}=1, N=4\cdot10^{6}$,)
which was dynamically evolved over 100 $N$-body timescales. Afterwards we rescaled the model to $R_{H}=1+\sqrt{2}$ in order to
compare it with the analytical velocity dispersion profile for anisotropic systems \citep{2002A&A...393..485B}. The usage of
high radial order terms ($n=30$) in combination with lowest order (spherical) angular terms ($l=0$) yields accurate results.} 
\label{sigma}
\end{figure}
The order of the radial expansion coefficient, $n$, required to guarantee the conservation of the initial space phase distribution
is investigated in more detail in this Section. For that purpose we evolved spherical Hernquist, Jaffe and S\'{e}rsic models forwards
in time up to 100 $N$-body timescales ($\Delta t_{\scriptsize{\mbox{up}}}=1$). Several different values for the radial expansion
coefficients, $n$, were chosen. Afterwards, we compared the numerical outcomes to the initial models in terms of density profiles, axis ratios
at several Lagrange radii (1,2,5,10,25,50,60,70,80,90\%), radial\footnote{The radial velocity dispersion, $\sigma_{r}^{2}$, is a sensible indicator
for model stability by tracing minuscule instabilities along $r$.} and tangential velocity dispersion as well as velocity
anisotropy parameter, $\beta(r)=\left(1+R_{A}^2/r^{2}\right)^{-1}$. All profiles were found to be perfectly stable when represented by high radial $n=30$ and
lowest angular order ($l=0$). This is in agreement with the results obtained in \S~\ref{subsec:testing_discreteness}. These are the
coefficients adopted for the main computations\footnote{The Hernquist base function of the SCF algorithm uses a different
scale length $a$, hence a high radial order is required for these models as well.}. For illustration the velocity dispersion
profile of an anisotropic Hernquist model with scale length $a=\left(1+\sqrt{2}\right)^{-1}$ is compared to the analytical
profile in Fig.~\ref{sigma}.

The expansion coefficient evaluation of axisymmetric and triaxial models is postponed.
Mesh effects and the dynamical friction routine (\S~\ref{subsubsec:df}) are investigated
and tested in the next Section \S~\ref{subsec:testing_grid}.

\subsection{Dynamical Friction and Grid Effects}\label{subsec:testing_grid}
\begin{figure}
\centering
\includegraphics[width=8.5cm]{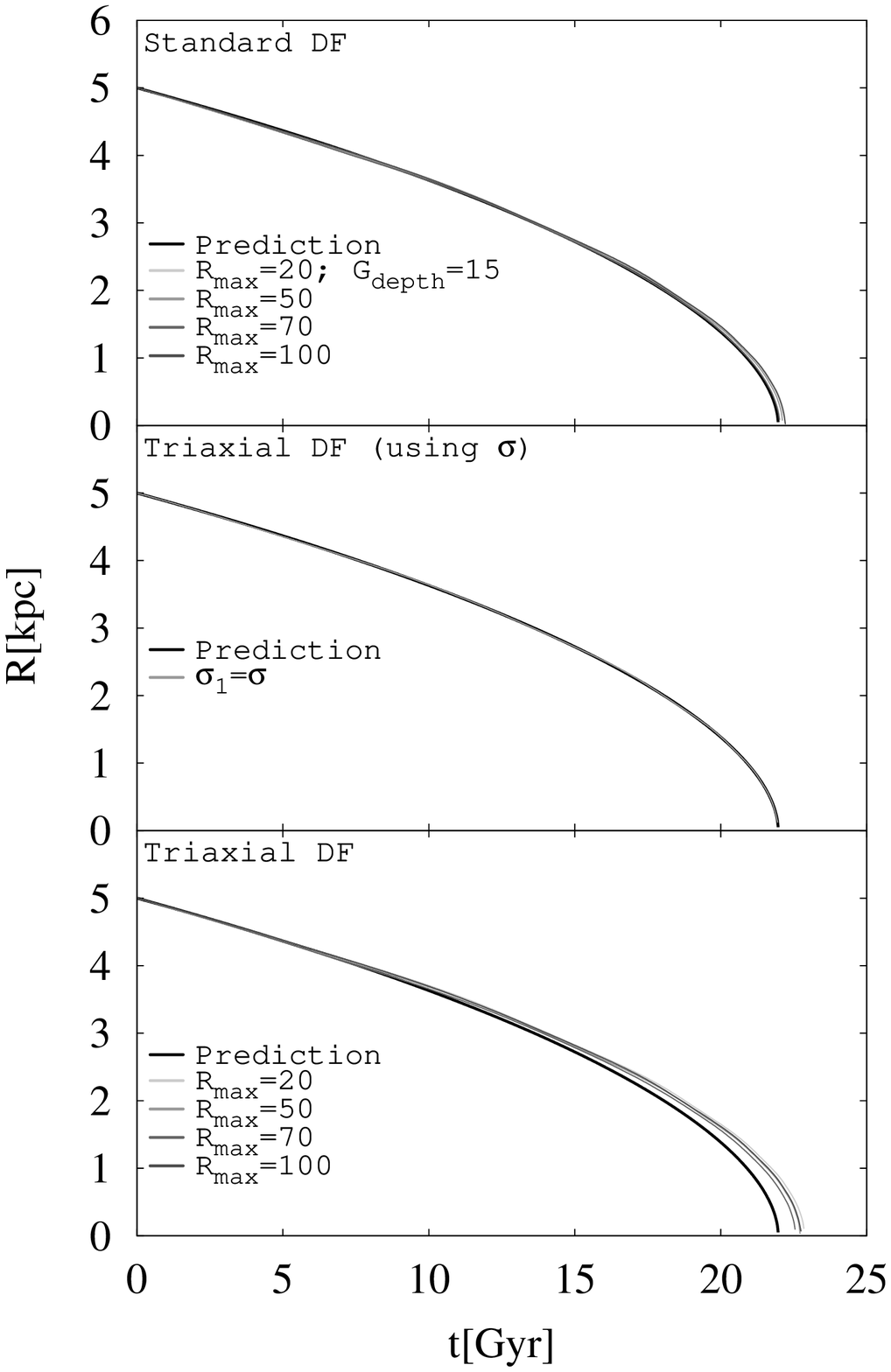}
\captionsetup{format=plain,labelsep=period,font={small}}
\caption{Decaying (circular) globular cluster ($m_{\scriptsize{\mbox{GC}}}=\unit{10^{7}}{\msun}$) orbit as a function of different grid
realizations and dynamical friction formulas. The standard DF computations match the predicted value within 1\% accuracy (top)
while the triaxial generalization formula yields slightly larger inspiral timescales $t_{\scriptsize{\mbox{fric}}}$ (bottom). 
The reason for this are discreteness noise effects in the evaluation of the largest eigenvalue $\sigma_{1}$ in combination with
the use of $\sigma_{1}^{3}$ in the denominator of Equation~\ref{f4}. By replacing $\sigma_{1}$ with the root mean square $\sigma$, which
ideally should be equal in isotropic models, the triaxial DF formula matches the analytical prediction (middle).
With regard to the difficulty of implementing highly nonlinear DF forces, a standard deviation of 1\% from the predicted
value, $t_{\scriptsize{\mbox{fric}}}\approx22$ Gyr, for Chandrasekhar's (standard) formula and 4\% for its triaxial generalization
is an excellent result. Using even larger particle numbers than $N=4\cdot10^{6}$ yields even more precise values.} 
\label{DF}
\end{figure}
The purpose of this Section is two-fold. The DF routine itself has to be tested and compared with analytical predictions of
idealized DF problems. Systematic effects caused by discreteness noise and grid selection effects have to be evaluated in
isotropic and anisotropic velocity distributions as well. In principle they can affect the computations and the calculus of
the velocity dispersion tensor which is required for the generalized dynamical friction force (Eq.~\ref{f3}). We numerically evaluated
the inspiral time $t_{\scriptsize{\mbox{fric}}}$ of a GC on a circular orbit ($r=\unit{5}{\mbox{kpc}}$) by using
Equation~\ref{f3} as well as Chandrasekhar's (standard) dynamical friction formula for a Maxwellian velocity distribution:
\begin{align}\label{f11}
\vec{a}_{\scriptsize{\mbox{GC}}}  = & -\frac{4\pi G^{2}m_{\scriptsize{\mbox{GC}}} \rho\left(\vec{r}\right)\ln\left(\Lambda \right)}
                                       {v_{\scriptsize{\mbox{GC}}}^{3}}\times  \\
                                       & \left[\mbox{erf}\left(X\right)-\frac{2X}{\sqrt(\pi)}\exp(-X^{2})
                                       \right]\vec{v}_{\scriptsize{\mbox{GC}}} \notag
\end{align}
Here $X=v_{\scriptsize{\mbox{GC}}}/(\sqrt{2}\sigma)$ and $\sigma=\sqrt{\left(\sigma_{1}^{2}+\sigma_{2}^{2}+\sigma_{3}^{2}\right)/3}$
is the one dimensional velocity dispersion obtained from the eigenvalues of the local velocity dispersion tensor.  
The numerical outcomes were then compared with analytical predictions of a decaying
($m_{\scriptsize{\mbox{GC}}}=\unit{10^{7}}{\msun}$) GC orbit and plotted in Fig.~\ref{DF}. The spherical galaxy had an
isotropic velocity distribution, Jaffe density profile \citep{1983MNRAS.202..995J}, total mass
$M_{\scriptsize{\mbox{GAL}}}=\unit{10^{11}}{\msun}$ and half mass radius $R_{H}=\unit{5}{\mbox{kpc}}$, similar to the
properties of NGC~4494 (\S~\ref{subsubsec:init_cond_gc_phasespace}). We selected a Jaffe profile because of its analytical
simplicity. Furthermore it deviates from the underlying base model of the SCF algorithm. In this way the reliance on 
higher order base functions was automatically tested as well. We fixed the Coulomb logarithm to $\ln{\Lambda}=6$ in order
to ease the analytical calculation. Grid selection effects were investigated by carrying out integrations with $N=4\cdot10^{6}$
particles and several realizations of the 5x5x5 grid (\S~\ref{subsubsec:df}). 
Evidently Chandrasekhar's (standard) dynamical friction formula, its triaxial generalization and the analytical prediction yield
(nearly) equal inspiral times $t_{\scriptsize{\mbox{insp}}}$ despite the highly nonlinear character of DF (Fig.~\ref{DF}). Owing to the 
IDW interpolation method (\S~\ref{subsubsec:df}), the chosen grid configuration does not have any visible systematic effect on $t_{\scriptsize{\mbox{insp}}}$.
And indeed, the differences in $t_{\scriptsize{\mbox{insp}}}$ caused by grid selection effects are only of the order of 200 million
years, compared to $\approx$20 Gyr absolute integration time, or 1\% in relative terms. Without the IDW interpolation method they
are much bigger (of the order of 25\%). The generalized triaxial DF routine (Eq.~\ref{f3}) yields slightly larger ($\approx$4\%) inspiral times
than the standard DF formula. The reason for this are discreteness noise fluctuations in the evaluation of the eigenvalues $\sigma_{i}$
in combination with the systematic use of $\sigma_{1}^{3}$ in the denominator of Equation~\ref{f4}. By replacing (for test purposes in
isotropic velocity distributions only) the denominator $\sigma_{1}^{3}$ with $\sigma^{3}$, the curve matches the analytical
prediction (Fig.~\ref{DF}). Nevertheless, in axisymmetric or triaxial galaxies, the anisotropic velocity distribution greatly overwhelms
discreteness noise fluctuations.

\begin{figure}
\centering
\includegraphics[width=8.5cm]{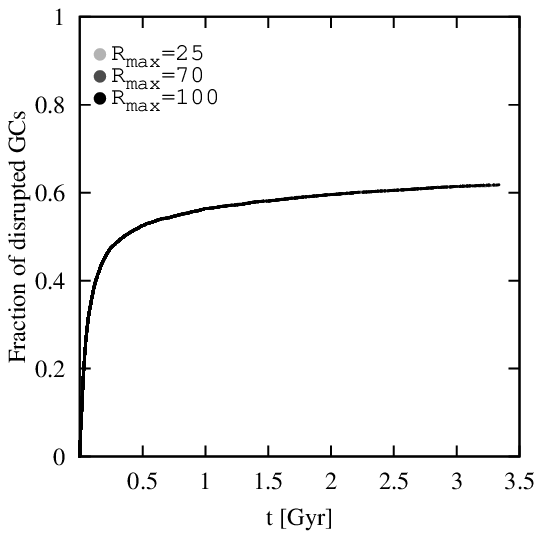}
\captionsetup{format=plain,labelsep=period,font={small}}
\caption{Owing to the IDW interpolation method, different grid configurations no longer have any discernible effects on the
cluster tidal disruption rate. $R_{\scriptsize{\mbox{max}}}$ is the outermost mesh size which affects the positioning of the inner grid cells.} 
\label{DFT}
\end{figure}

To further explore mesh and discreteness noise effects in anisotropic velocity distributions, we performed a second test.
The orbits of 10,000 globular clusters in a strongly triaxial galaxy with a cored density profile and a radially biased velocity
distribution at large distances ($\beta \approx 1$) were evolved forward in time under the influence of the generalized dynamical
friction force (Eq.~\ref{f3}). The generation of this particular model is described in \S~\ref{subsubsec:init_cond_gc_phasespace}.
Total mass and scale of the galaxy were assumed to be identical to the test before. For this test we 
restrict to tidal disruption processes from Eq.~\ref{f1} and neglect the long term energy-equipartition driven evaporation.
The cluster masses and sizes were distributed according to the single power
law mass spectrum (Eq.~\ref{f5}) and present day half mass relation (Eq.~\ref{f10}). Discreteness noise and mesh effects were artificially enhanced
by computing a small number $N=5\cdot10^{5}$ model over 200 $N$-body timescales ($\approx 3.5$ Gyr). Moreover, the overall potential
was frequently updated by the SCF algorithm in $\Delta t_{\scriptsize{\mbox{up}}}=1$ intervals. We then compared the timescales until
given fractions of GCs were destroyed in computations with identical physical properties apart from different grid size realizations.
In this way mesh effects were isolated and analyzed. The numerical outcomes are shown in Figure~\ref{DFT}. Evidently, the IDW
interpolation method narrows grid effects down to insignificant values. The implemented DF routines yield credible results
with insignificant errors. 

\subsection{Potential Fluctuations and their Relevance for Globular Cluster Disruption}\label{subsec:testing_fluctuation}

Finally, the performance of the disruption routine (\S~\ref{subsubsec:tidal_shocks}) has to be evaluated. Notwithstanding that the base
functions as well as their derivatives are continuous, summing up these functions might lead to wiggles along the radial direction.
It may therefore be possible that the radial acceleration $a_{r}(r)$ as well as the evaluation of the Jacobi radius (Eq.~\ref{f1}) are affected 
and a cluster is incorrectly assumed to be destroyed by tidal forces.
To guarantee computational outcomes free of biased GC disruption rates, we performed several test integrations by considering tidal
disruption processes only, i.e. by neglecting the long term relaxation driven mass loss. Isotropic Hernquist
models ($N=10^6, n=30, l=0,\Delta t_{\scriptsize{\mbox{up}}}=1$) with 5,000 randomly distributed clusters were evolved forward in time.
They were scaled to the physical properties of MOD1 (M~32) and MOD4 (NGC~4889, Table~\ref{tablen=1}), the two galaxies in our sample with the most
extreme GC-to-galaxy half mass ratios, $r_{H}/R_{H}$. For each of these models the Jacobian radius was evaluated directly from the SCF
algorithm as well as by using an analytical expression for the radial acceleration. The differences in the total number of disrupted clusters
$N_{\scriptsize{\mbox{dis}}}$ were below 0.05\% (M~32) and 0.3\% (NGC~4889). This corresponds to absolute discrepancies of two and one
globular cluster(s) respectively. Being so small in magnitude, these fluctuations can be neglected, particularly because the main
computations are performed with twenty times higher particle numbers and without re-evaluation of the potential.

\section{Extreme radial anisotropy}\label{ERA}
\begin{figure}
\centering
\includegraphics[width=8.5cm]{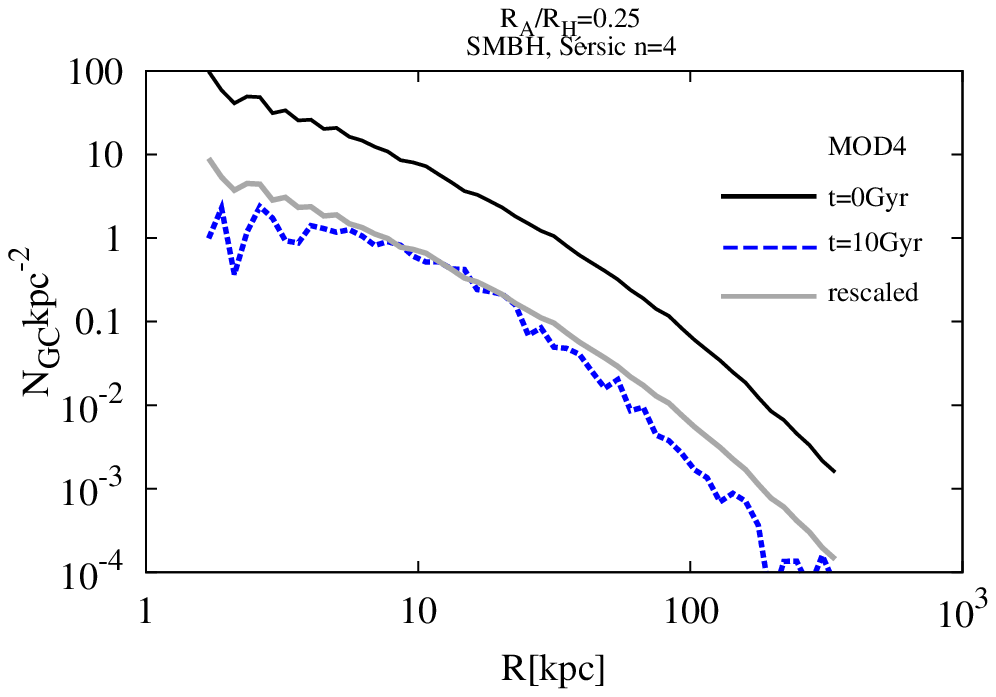}
\captionsetup{format=plain,labelsep=period,font={small}}
\caption{Number density profile of a radially biased S\'{e}rsic n=4 model with extreme anisotropy, $R_{A}/R_{H}=0.25$. At some point, the slope becomes steeper than that of the 
initial distribution. The gray line corresponds to the rescaled initial profile. The scaling factor is arbitrary and only serves for comparison issues.} 
\label{denN4889ra025}
\end{figure}

\begin{figure*}
\centering
\includegraphics[width=16cm]{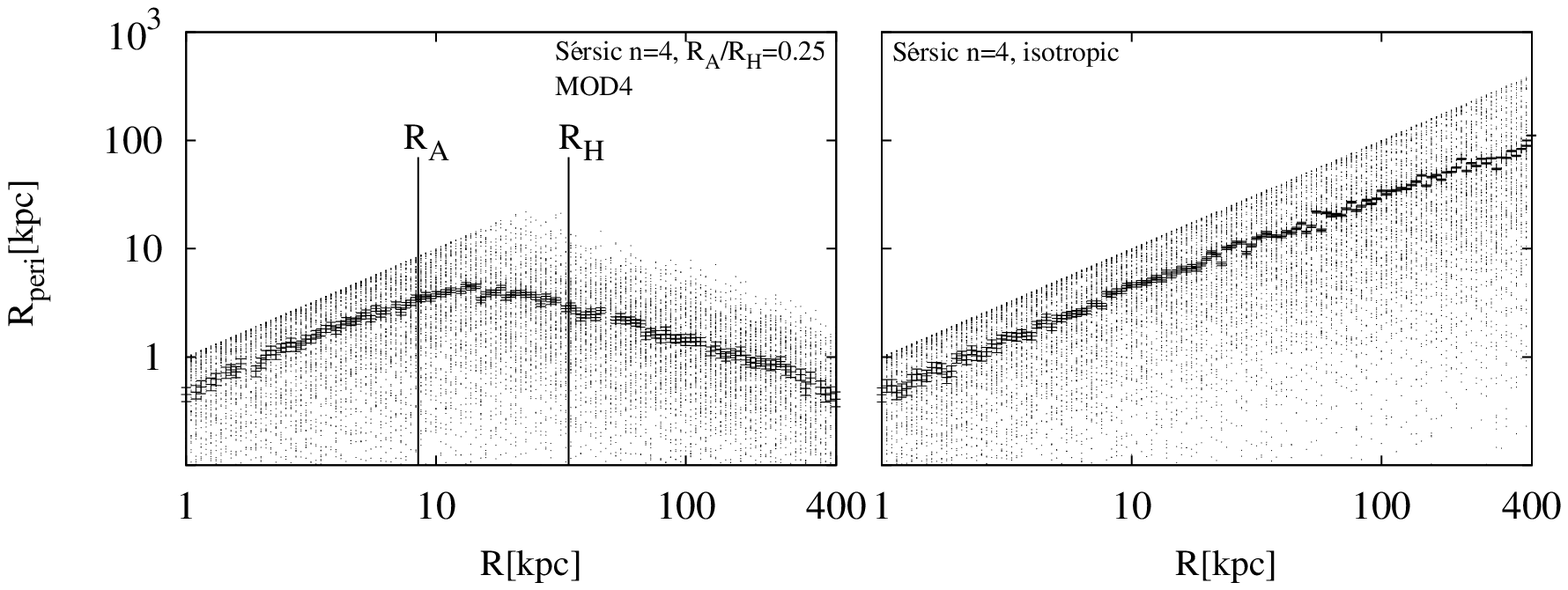}
\captionsetup{format=plain,labelsep=period,font={small}}
\caption{Monte-Carlo realizations of pericentre versus galactocentric distances of reference model MOD4 with a S\'{e}rsic n=4 density profile, 
central SMBH and a strongly radially biased (left panels) and isotropic (right panels) velocity distribution. The central (black) data points
represents the arithmetic mean. A truncation in the strongly radially biased configuration is evident. It explains why the outer GC profiles
in the $R_{A}/R_{H}=0.25$ models are more strongly influenced beyond a characteristic radius being of the order of $R_{A}=8.5$ kpc.} 
\label{MC}
\end{figure*}

The final number density profile of a radially biased S\'{e}rsic n=4 model with extreme anisotropy, $R_{A}/R_{H}=0.25$, is plotted in Fig.~\ref{denN4889ra025}. The profile shows
a less pronounced core than the corresponding isotropic configuration (Fig~\ref{GCProfiles}). The reason for this is that clusters all along the spatial extent of the galaxy are eroded efficiently.
However, at very large galactocentric distances, the slope becomes steeper. This is due to the fact that pericenter distances at a given 
galactocentric distance start to decrease again. This effect is illustrated by means of independent Monte-Carlo Computations in Fig.~\ref{MC}.
Although there are reports on galaxies in which the globular
cluster system profiles at large radii are indeed steeper than the surface brightness profile of their host galaxies, e.g. NGC~4406 (M~86)
\citep{2004AJ....127..302R, 2009A&A...507..183C}, such a highly radially biased (initial) GC configuration does not represent a plausible
explanation, especially as the degree of radial anisotropy remains still very large at huge galactocentric distances.
In the case of M~86 it was suggested by \cite{2004AJ....127..302R} that tidal truncation induced by M~84
or even the potential well of the Virgo cluster is responsible for the steep profile. The much flatter outer GCS profile of its companion
M~84 might support this scenario \citep{2009A&A...507..183C}.

\label{lastpage}
\end{document}